\documentclass[10pt,journal,compsoc]{IEEEtran}

\ifCLASSOPTIONcompsoc
  \usepackage[nocompress]{cite}
\else
  \usepackage{cite}
\fi
\ifCLASSINFOpdf
\else
\fi
\usepackage{mathptmx} 
\newcommand{\ignore}[1]{}
\usepackage{afterpage}
\usepackage{enumitem}
\usepackage[normalem]{ulem}
\usepackage[hyphens]{url}
\usepackage{hyperref}
\usepackage{enumitem}
\usepackage{graphicx}
\usepackage{soul}
\usepackage{color}
\usepackage{subfig}
\usepackage[fleqn]{amsmath}
\usepackage{relsize}
\usepackage{algorithm}
\usepackage[noend]{algpseudocode}
\usepackage{hhline}
\newcommand{\specialcell}[2][c]{%
  \begin{tabular}[#1]{@{}c@{}}#2\end{tabular}}

\begin{document}
%
\title{ReCA: an Efficient Reconfigurable Cache Architecture for Storage Systems with Online Workload Characterization}
%
%
%
%


\author{Reza~Salkhordeh,
        Shahriar~Ebrahimi,
        and~Hossein~Asadi,~\IEEEmembership{Senior Member,~IEEE}
\IEEEcompsocitemizethanks{\IEEEcompsocthanksitem Reza Salkhordeh, Shahriar Ebrahimi, and Hossein Asadi are with the Department
of Computer Engineering, Sharif University of Technology, Tehran,
Iran.\protect\\
E-mail: {salkhordeh,shebrahimi}@ce.sharif.edu and asadi@sharif.edu
}}

%
%


\markboth{}%
{Salkhordeh \MakeLowercase{\textit{et al.}}: ReCA: an Efficient Reconfigurable Cache
Architecture for Storage Systems with Online
Workload Characterization}
%



\IEEEtitleabstractindextext{%
\begin{abstract}
In recent years, \emph{Solid-State Drives} (SSDs) have gained tremendous attention in computing and storage systems due to significant performance improvement over \emph{Hard Disk Drives} (HDDs).
The cost per capacity of SSDs, however, prevents them from entirely replacing HDDs in such systems.
One approach to effectively take advantage of SSDs is to use them as a caching layer to store performance critical data blocks in order to reduce the number of accesses to HDD-based disk subsystem.
Due to characteristics of Flash-based SSDs such as limited write endurance and long latency on write operations, employing caching algorithms at the \emph{Operating System} (OS) level necessitates to take such characteristics into consideration.
Previous OS-level caching techniques are optimized towards only one type of application, which affects both generality and applicability.
In addition, they are not adaptive when the workload pattern changes over time.
This paper presents an efficient \emph{Reconfigurable Cache Architecture} (ReCA) for storage systems using a comprehensive workload characterization to find an optimal cache configuration for I/O intensive applications.
For this purpose, we first investigate various types of I/O workloads and classify them into five major classes.
Based on this characterization, an optimal cache configuration is presented for each class of workloads.
Then, using the main features of each class, we continuously monitor the characteristics of an application during system runtime and the cache organization is reconfigured if the application changes from one class to another class of workloads.
The cache reconfiguration is done online and workload classes can be extended to emerging I/O workloads in order to maintain its efficiency with the characteristics of I/O requests.
Experimental results obtained by implementing ReCA in a 4U rackmount server with SATA 6Gb/s disk interfaces running Linux 3.17.0 show that the proposed architecture improves performance and lifetime up to 24\% and 33\%, respectively.
\end{abstract}

}

\maketitle


\IEEEdisplaynontitleabstractindextext

%
\IEEEpeerreviewmaketitle

\ifCLASSOPTIONcompsoc
\IEEEraisesectionheading{\section{Introduction}\label{sec:introduction}}
\else
\section{Introduction}
\label{sec:introduction}
\fi
\IEEEPARstart{W}{ith} the ever-increasing demand for computational power, applications have become orders of magnitude more performance consuming compared with their descendants.
To meet the performance demands, the computational power of computer systems has been increased by orders of magnitude in the recent decades.
Storage systems, on the other hand, had a marginal performance improvement in the same timespan.
\emph{Hard Disk Drives} (HDDs) as the conventional devices employed in storage systems, have mechanical components which puts a tight upper limit on their maximum performance.
This has made storage systems as a major performance bottleneck in computer systems.


\begin{figure}
\centering
\includegraphics[scale=0.27]{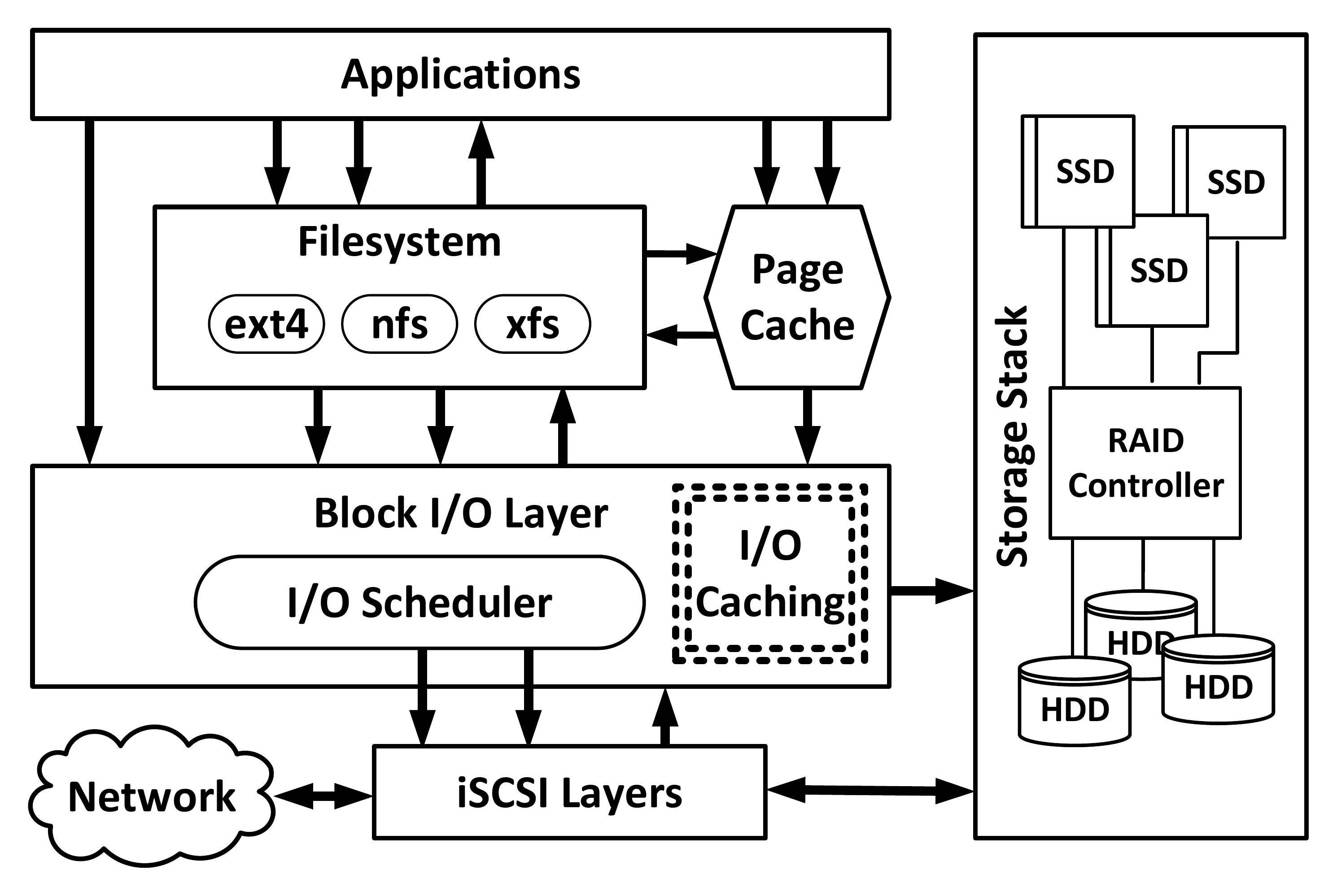}
\vspace{-.3cm}
\caption{Linux I/O stack diagram}
\label{fig:storagestack}
\vspace{-.6cm}
\end{figure}

To alleviate the performance gap between storage systems and the other parts of computer systems, many approaches have been employed at the \emph{Operating System} (OS)  level in recent years.
Optimizing the ordering of I/O requests \cite{axboe2007cfq}, placing relevant data pages close to each other \cite{ts2002planned}, and merging small requests \cite{axboe2004linux} are some of the basic approaches proposed at the OS-level aimed to enhance performance of storage systems.
Such optimizations are mostly implemented at the OS block I/O layer since it routes all I/O requests issued from both applications and filesystems to the disk subsystem and has necessary means to optimize the storage system (shown in Fig. \ref{fig:storagestack}).
In addition to optimizing the requests in the block I/O layer, employing a high performance device as a caching layer for HDDs has been practiced in recent years.
As an example, the free space in the main memory can be used as a cache for storage devices in order to reduce the average response time.
Due to the volatile characteristic of the main memory, this technique cannot be reliably used in storage systems, where data loss due to power outage result in significant customer dissatisfaction.

\begin{figure}
\centering
\includegraphics[scale=0.65]{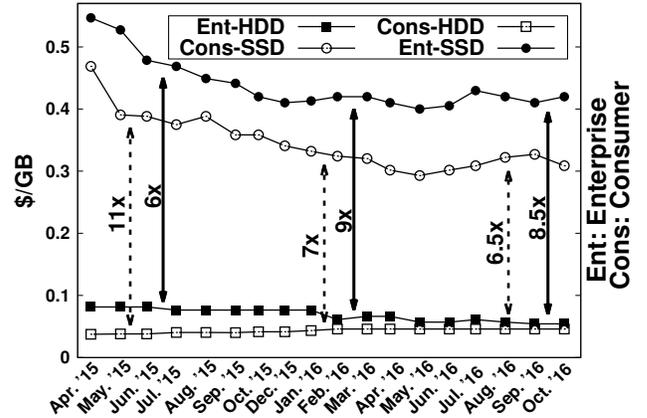}
\vspace{-.2cm}
\caption{SDD and HDD Price Gap Trend}
\label{fig:ssdprice}
\vspace{-.5cm}
\end{figure}

To cope with both limited performance of HDDs and data loss concern in the main memory, recent studies have suggested using \emph{Solid-State Drives} (SSDs) as a caching layer for HDDs \cite{Pritchett:2010:SHE:1815961.1815982,Saxena:2014:DPS:2661087.2629491} (a.k.a., I/O caching shown in Fig. \ref{fig:storagestack}).
Enterprise SSDs have 8.5x higher cost compared to enterprise HDDs and the price gap between them will not be closed in the upcoming years as the price trend suggests \cite{ssdprice,pcpartpicker} (shown in Fig. \ref{fig:ssdprice}). 
Therefore, SSDs should be used alongside with HDDs in order to design a cost-efficient, high-performance storage system.
To effectively use SSDs as the OS cache layer, modifications in the conventional caching policies such as \emph{Least Recently Used} (LRU) and Clock have been suggested by the previous studies \cite{1297303,Huang:2016:IFD:2888404.2737832,Santana:2015:AA:2813749.2813763}.
Another suggested approach in the previous work is prioritizing various request types such as filesystem metadata, read, and random requests by using either absolute or relative priority \cite{Azor,Hystor,Mesnier:2011:DSS:2043556.2043563,hstoragedb}.
The absolute priority employed in previous studies is inefficient as observed by a recent study \cite{tiering}.
On the other hand, while the relative priority has more flexibility, the complexity of the request types and the storage devices (in particular SSDs) makes this approach inefficient in many cases.
In addition, the coefficients for the relative priority cannot be adjusted dynamically in case a workload pattern changes during runtime.
Another group of previous studies attempts to improve the LRU-based algorithms used as the eviction policy in the caching solutions \cite{Huang:2016:IFD:2888404.2737832,Santana:2015:AA:2813749.2813763,7255203,1297303}.
Unlike caching in the memory subsystem, improving the hit ratio in the storage system will not necessarily improve the average response time.
This can be justified by the fact that a caching algorithm with a lower hit ratio that mostly redirect sequential requests to HDDs will likely provide higher performance than a caching algorithm with higher hit ratio that mostly redirect random requests to HDDs.
This is due to HDDs demonstrate at least two orders of magnitude higher performance in sequential requests compared to random requests.

This paper proposes a \emph{Reconfigurable Cache Architecture} for storage systems, called \emph{ReCA}, which aims to improve system performance by categorizing I/O intensive applications and their characteristics with respect to their performance on various cache configurations.
Based on a comprehensive characterization on the performance of various request types for both HDDs and SSDs, ReCA classifies I/O intensive applications into five major categories: a) random consumers, b) sequential producer/consumers, c) random producer/consumers, d) archival consumers, and e) large file generators.
The proposed characterization study leads us to design an optimal cache configuration for each category of workloads to maximize system performance.
Based on the observed characteristics of the running workload, one of the five categories which suits best the workload will be dynamically adjusted.
Emerging workload types can be easily added to ReCA by modifying a data file without need for changing the architecture and/or re-compiling the code.
This can also be done online during runtime.
ReCA is also compatible to run multiple applications simultaneously by characterizing each application separately.
ReCA will be reconfigured as soon as the running workload pattern transforms into another workload category.
To the best of our knowledge, none of the previous studies have suggested a reconfigurable cache architecture for storage systems taking into account the workload patterns to improve I/O performance.

We have implemented ReCA on an open source caching platform, called \emph{EnhanceIO} \cite{enhanceio}, with addition of more than 500 lines of code to the OS block I/O layer.
The experiments have been done using a 4U server with SATA 6 Gb/s HDDs and SSDs to remove the interface bottlenecks.
Linux kernel 3.17.0 is used throughout the experiments as the OS kernel of the testbed.
The workloads used in the experiments consists of more than 75 traces from FileBench \cite{filebench}, FIO \cite{fio}, Postmark \cite{postmark}, HammerDB \cite{hammerDB}, and many other publicly available storage traces.
Experimental results show that ReCA can improve performance up to 24\% (16\%, on average) for various workloads compared to previous studies while removing the need for mirrored-cache configuration for SSDs in many applications and improving the lifetime of SSDs up to 33\%.

The main contributions of this work are as follows.
\begin{itemize}[leftmargin=*]
\item By examining performance of various storage devices under several request types, we demonstrate that assumptions made by
previous studies on SSD performance such as superior read performance of SSDs compared
to its write performance \cite{5366737,Saxena:2012:FLC:2168836.2168863} may not hold in all workloads or request types.
\item A comprehensive workload characterization over 75 traces has been conducted in this paper to analyze I/O behavior of diverse applications with respect to various SSD caching policies and architectures which has not been done in previous studies.
\item We conducted a detailed analysis investigating the effect of filesystem metadata on performance of storage systems which demonstrates that \emph{(a)} filesystem metadata requests do not have the same request type (as opposed to the assumptions made in previous studies) and \emph{(b)} the impact of filesystem metadata requests significantly differs across workload categories.
Such important observations are fed into the proposed caching architecture to further improve performance of storage systems.
\item Based on the proposed workload characterization, a \emph{Reconfigurable Cache Architecture} (ReCA) has been proposed which aims to dynamically adapt to the changes in the I/O behavior and reconfigures itself toward optimal cache policy and configuration for the currently running workload.
Emerging workload categories can be added to ReCA in runtime to be able to optimize itself toward new workloads.
ReCA is able to reconfigure \emph{cache line size}, \emph{write policy}, and \emph{eviction policy} online without any prior knowledge on the pattern of the running workload.
This is achieved by an online monitoring system designed into ReCA that can adapt to the workload changes.
\item While ReCA mainly aims to improve performance by optimizing cache architecture, it also removes the need for redundant SSDs in four out of five workload categories by employing \emph{write-through} policy.
This policy selection results in reducing the cost of caching (by removing redundant SSDs) and improving reliability by not keeping dirty data pages in the I/O cache.
In addition, the lifetime of SSDs is improved (up to 33\%) in two workload categories by redirecting all write requests to the disk subsystem.
\end{itemize}

The rest of this paper is organized as follows.
Section \ref{sec:related} discusses the related work.
In Section \ref{sec:char}, an analysis on the workload characterization will be presented.
The proposed architecture is introduced in Section \ref{sec:proposed}.
The overall workflow of the proposed architecture is detailed in Section \ref{sec:workflow}.
The experimental setup and results are reported in Section \ref{sec:results}.
Finally, Section \ref{sec:conclusion} concludes the paper.

\begin{figure}[!t]
\centering
\subfloat[Tiering]{\includegraphics[width=.24\textwidth]{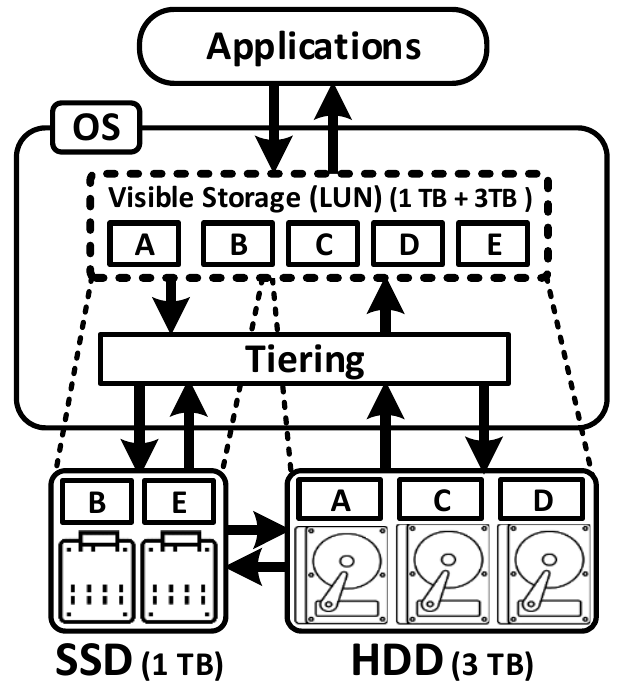}%
 \label{fig:tiering}}
\hfil
\subfloat[Caching]{\includegraphics[width=.22\textwidth]{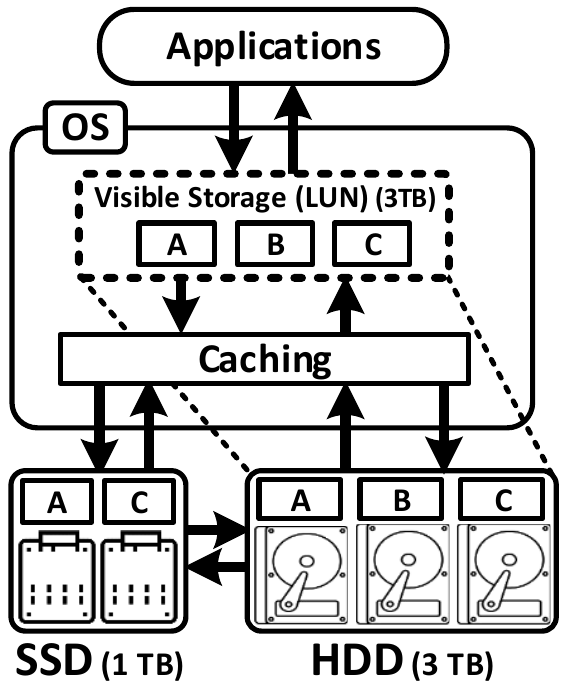}%
\label{fig:caching}}
\vspace{-.2cm}
\caption{Various Storage Architectures using SSDs}
\label{fig:cachingtiering}
\vspace{-.55cm}
\end{figure}

\vspace{-.3cm}
\section{Related Work}
\label{sec:related}
 SSDs can be integrated into the storage stack as a caching layer or as a high performance tier above HDDs in a tiering technique as depicted in Fig. \ref{fig:cachingtiering}.
Both of these techniques have been examined in the previous studies and many algorithms have been proposed using either tiering or caching techniques.
Tiering attempts to split data pages between storage devices called \emph{tiers} in order to reach the required performance, power, and cost efficiency \cite{tiering,fsmac,5581609} (shown in Fig. \ref{fig:tiering}).
In tiering, a data page will only reside on one of the storage devices, while in caching techniques (Fig. \ref{fig:caching}), a copy of data block will be moved to the cache.
Hence, the space efficiency of tiering is higher than caching, as illustrated in Fig. \ref{fig:cachingtiering}.
Tiering solutions mainly focus on migrating data pages between tiers and are mostly suitable for steady workloads without any sudden changes in workload pattern \cite{storagestack}.

Caching solutions, on the other hand, try to adapt themselves with the sudden changes in the workload pattern and perform more efficient in workloads with higher locality \cite{storagestack}.
In addition, they are suitable for computer systems running many applications simultaneously or running virtualization platform \cite{Meng:2014:VAS:2643634.2643649}.
LRU is used as the baseline algorithm for most caching solutions and many improvements for this algorithm have been suggested in previous studies.
A technique based on \emph{set sampling} and \emph{set dueling} has been suggested in \cite{Qureshi:2007:AIP:1250662.1250709} to
choose between two eviction policies for a group of cache sets based on hit ratio. Despite this technique is the
best metric for performance evaluation of CPU caches, it is not suitable for SSD caches due to their characteristics and endurance
limitations.
ARC \cite{1297303} maintains two LRU queues for capturing both recency and frequency of accesses to data pages to identify the most effective data pages for caching.
mARC \cite{Santana:2015:AA:2813749.2813763} is an extension of ARC which tries to identify unstable workloads and prevents caching data pages which might result in cache pollution while allowing performance critical data pages to be cached at the block I/O layer.
One of the drawbacks of ARC is neglecting the cost of placing data pages in cache for each missed access.
This limitation is addressed in LARC \cite{Huang:2016:IFD:2888404.2737832} by preventing cache pollution and keeping hot data pages in cache for a longer time.
LARC, however, reacts very slowly to changes in the workload pattern and misses many opportunities for caching performance critical data pages.

Another group of the previous studies have focused on prioritizing one type over other request types in order to exploit SSDs or workload characteristics.
AutoRAID \cite{Wilkes:1996:HAH:225535.225539} employs a redundant array (similar to RAID-1) as a tier for hot data pages
while storing cold data pages in RAID-5 array.
Hot data pages are identified dynamically which enables AutoRAID to reconfigure itself in case of workload changes.
HDD-based caching has also been suggested in \cite{Miranda:2014:COR:2591305.2591319,Bhadkamkar:2009:BBS:1525908.1525922} by dedicating a small portion of HDDs to hot data pages to reduce the disk head movement.
Although AutoRAID and a few of follow-up studies have reconfiguration ability, they mainly are HDD-based which is not applicable to emerging storage devices and applications (unlike SSD-based architectures such as ReCA).
A simple read/write counter for prioritizing requests has been suggested in \cite{5366764}.
Assigning absolute priority to metadata requests over regular requests has been suggested in \cite{Azor,macss,Hystor} which is demonstrated to be inefficient in many cases \cite{tiering}.
In addition, this approach requires modifications in several layers of OS which reduces the generality of this approach and is restricted to the examined filesystem since they use different algorithms and, moreover, modifications cannot be easily applied to the other filesystems.
SSDs have a relatively high performance on random requests, therefore, previous studies suggested assigning higher priority to random requests over sequential requests.
One technique that employs this prioritization was proposed in \cite{Hystor} where the priority for caching a data page is proportional to the inverse of its size.
A three-level table has been employed in order to calculate the overall priority of each data page and based on the priorities, candidate data pages will be selected for caching in SSDs.
A similar approach has been employed in \cite{Kang:2014:HSA:2660459.2660493} for improving performance in the virtualization environments.
RPAC \cite{7255203} is another approach that considers locality of accesses to HDDs.
By prioritizing data pages near each other and caching these data pages in SSDs, the locality of accesses in HDDs increases which will improve its performance.
HybridStore \cite{hybridstore} proposes a planner which optimizes the total cost of a storage subsystem based on
	its performance behavior and SSD lifetime using \emph{Integer Linear Programming} (ILP).
	In addition, a dynamic controller has been proposed that models performance behavior of SSDs in runtime.
	Unlike HybridStore, ReCA employs a more detailed workload characterization over real hardware and aims to optimize
	itself based on SSD characteristics.
	The dynamic controller of HybridStore, however, can be employed in ReCA to further increase its ability to
	enhance SSD lifetime.
Moirai \cite{Stefanovici:2015:SCM:2806777.2806933} is a hypervisor caching architecture able to provide various caching \emph{Quality-of-Service} (QoS) options such as minimum bandwidth per \emph{Virtual Machine} (VM) and employing different cache policies for VMs.
The dynamic approach of ReCA can be exported to hypervisor-based caching architectures such as Moirai \cite{Stefanovici:2015:SCM:2806777.2806933} and ECI-Cache \cite{ahmadian:eci} to further optimize caching for VMs.

The effect of workload characteristics on SSD failures has been investigated in \cite{ahmadian:date}.
LeavO cache \cite{Lee:2015:SCO:2695664.2695886} tries to improve reliability of SSD caching by keeping both old and new
data pages in SSD.
A more space-efficient architecture has been proposed in \cite{7573841} to reduce the number of writes to SSD cache which improves its lifetime by storing modifications compared to old data pages.
Employing a log-structured approach can also improve SSD lifetime as suggested in \cite{Min:2012:SRW:2208461.2208473,Mao:2012:HHP:2093139.2093143}.
Such architectures can be used alongside ReCA to further improve its endurance and/or reliability.

Modifying the interface between OS and the disk subsystem to further improve the performance is examined in \cite{Saxena:2014:DPS:2661087.2629491}.
Such techniques, however, require hardware and firmware modifications which limits both their applicability and generality.
Altering the storage stack to feed semantic information of applications to the caching layer has been examined in \cite{Mesnier:2011:DSS:2043556.2043563} to improve the performance of databases.
Such technique also suffers from poor generality.
SSD caching can also be employed in the client-server solutions \cite{180727,6232368,7266934}.
Reducing the number of writes in SSDs and improving their lifetime have been studied in many works \cite{7035001,CostEffective}.
Clustered tiering consists of HDD, SSD, \emph{Network Attached Storage} (NAS), and other storage devices has been suggested by several previous studies \cite{Kakoulli:2017:ODF:3035918.3064023,Oh:2017:TAM:3078468.3078485,Abd-El-Malek:2005:UMV:1251028.1251033}.
The main aim of these studies, however, differs from the goal of this paper which is mainly improving the performance of the storage subsystem.

Due to the importance of the request classification, several classification studies have also been proposed in previous work \cite{7155523,4636097,Roselli,Tarihi:2015:DAD:2745844.2745856}. The studies  presented in \cite{4636097,Roselli} are focused to investigate and 
characterize I/O requests using the semantic information available at the filesystem layer which vastly differs from the target information in our study at the block I/O layer.
Another characterization study was proposed in  \cite{7155523}, which aims to classify I/O requests at 
the storage device level to address long write latency, high write energy, and limited endurance of \emph{Phase-Change Memory} (PCM) cache designed for SSDs, by employing a small DRAM alongside PCM, while maintaining the same performance level of previous architectures.
The discussion of such studies targeting internal architecture of storage devices is beyond the scope of this paper. 

\begin{figure}
	\centering
	\includegraphics[scale=0.38]{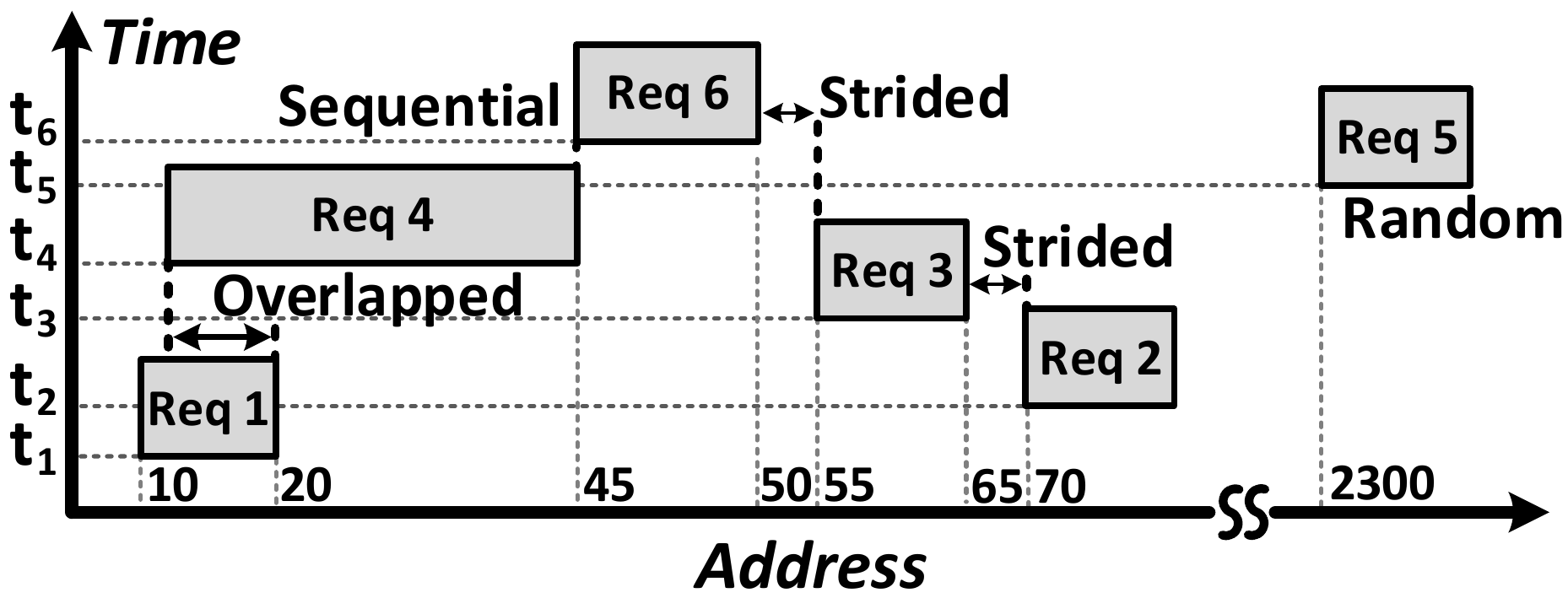}
	\vspace{-.23cm}
	\caption{Arbitrary sequence of requests over time}
	\label{fig:char}
	\vspace{-.6cm}
\end{figure}

\vspace{-.3cm}
\section{Workload Characterization}
\label{sec:char}
The main goal of any SSD caching architecture is to identify the performance critical data pages and predict the workload pattern.
The actual response time of requests in HDDs and SSDs significantly depends on the previous requests due to the HDD head position and internal buffers/caches in both HDD and SSD which prevents us from using average response time reported in device datasheets.
Therefore, before proposing an effective caching architecture, conducting a study on the workload patterns and methods for identifying performance critical data pages is crucial.
Here, we present a comprehensive study on identified parameters affecting the I/O performance which will be used in the proposed reconfigurable cache architecture in order to improve performance compared to the previous studies.

Since HDDs and SSDs have distinct characteristics and internal structures, they exhibit differing behavior over various request types.
On the other hand, analyzing the performance of the storage devices, which is crucial for choosing the right type of data pages for caching, highly depends on the request types.
Therefore, before analyzing the performance of storage devices, a classification on the request types is required.
This classification (Sec. \ref{sec:reqclass}) enables us to benchmark storage devices from various perspectives.
The complexity of the storage devices, specially SSDs, increases the importance of this benchmarking.
In addition, commercial SSDs employ many optimizations and buffering techniques in order to improve the performance and/or lifetime.
Moreover, the complex structure of SSDs cannot be predicted easily without such experiments.
Request interleaving is another important factor affecting the performance of the storage subsystem which will be discussed in detail in Sec. \ref{sec:interleaving}.
After careful analysis of the request types, workloads will be categorized and based on the proposed categorization, optimal cache configuration for each workload category will be presented in Sec. \ref{sec:workloadchar}.
Apart from the request types, there are many semantic information available at the OS level which can help us identify the request pattern for each data page.
Requests issued for accessing filesystem metadata are one of the most important data requests as they have significant effect on the filesystem performance \cite{tiering,Hystor,Azor}.
As such, a complete analysis has been done in this work regarding filesystem metadata and how a caching architecture can benefit from such analysis (Sec. \ref{sec:metadata}).

\vspace{-.2cm}
\subsection{Request Classifications}
\label{sec:reqclass}
In the proposed classification, I/O requests are classified into four major classes which are: a) \emph{sequential}, b) \emph{random}, c) \emph{strided}, and d) \emph{overlapped}.
Since the response time of a request is dependent to the previous requests, each request will be evaluated and classified based on the previous requests.
Fig. \ref{fig:char} depicts an arbitrary sequence of requests over time.
Upon arrival of a request, it will be compared to the previous requests in order to identify its type.
The number of the past requests used for comparison depends on the general architecture of the OS and employed devices.
Here, we assume size of 64 requests for request history queue which is close to the queues employed in the disk and the I/O scheduler in OS.
In Fig. \ref{fig:char}, request 6 is classified as \emph{sequential} since its starting point is exactly after the ending point of one of the requests issued earlier.
The starting point of a \emph{strided} request has a small gap from the ending point of one of the previous requests similar to request 2 compared to request 3 in Fig. \ref{fig:char}.
The performance of the strided requests is lower than the sequential requests since in rotational disks, disk head has to move a few sectors to start responding to the second request.
In addition, I/O scheduler in OS tries to merge sequential requests before sending them to disk and strided requests cannot be merged together because of the existing small gap between them.
If there is no request adjacent to a request, similar to request 5, it will be considered as \emph{random} which will have the worst performance on HDDs compared to other request types.
Request 4 is an \emph{overlapped} request since it shares a data page with one of the previous requests.

\begin{table}[t]
	\caption{Testbed configuration}
	\vspace{-.3cm}
	\label{tbl:config}
	\scriptsize
	\centering
	\begin{tabular}{|c|c|c|c|}
		\hline
		Device & Model & \specialcell{Sequential\\Read/Write*} & \specialcell{Random\\Read/Write*}  \\ \hline
		CPU &  Xeon E5620 & - & - \\  \hline
		Memory &  32 GB DDR3 & \multicolumn{2}{|c|}{10 GB/s} \\ \hline
		HDD & Western Digital Red Pro HDD  & 120 MB/s & 80/150 IOPS  \\     \hline
		SSD &   Samsung SSD 850 PRO & 480/440 MB/s & 20k/30k IOPS \\    \hline
	\end{tabular}
	*Actual measured performance in our experiments
	\vspace{-.5cm}
\end{table}


In order to obtain the performance characteristics of SSDs and HDDs, we have benchmarked two disk drives with the request classes using \emph{IOMeter} benchmarking tool \cite{iometer1997iometer}.
The specification of the drives used for this experiment is presented in Table \ref{tbl:config}.
Since our goal is to measure the overall disk performance, all disk optimizations such as read-ahead or disk cache have been enabled.
Random requests are 4KB requests where their addresses are randomly distributed over the address space.
The probability distribution for choosing the request addresses was set to uniform distribution to measure the pure random performance of disks.
The sequential requests have 4KB size and the address of each request is set to the ending address of the last request.
Strided requests have the same size as sequential requests and each request is issued to the ending address of the last request, plus an offset which is set to 8KB.

\begin{figure}
	\centering
	\includegraphics[scale=0.65]{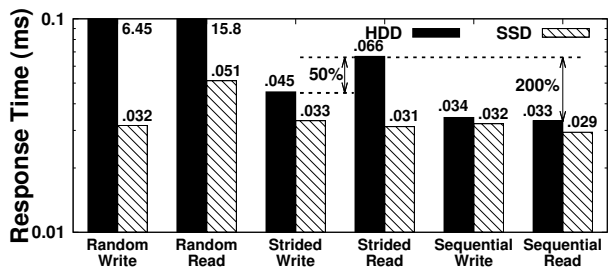}
	\vspace{-.2cm}
	\caption{HDD and SSD latencies for various request types}
	\label{fig:charperf}
	\vspace{-.7cm}
\end{figure}

Fig. \ref{fig:charperf} shows the average response time of the benchmarked SSDs and HDDs for the mentioned request classes.
Read and write requests are separated since HDDs and SSDs demonstrate different performance behavior.
In addition, the embedded cache in disks can cache all write requests (when running in write-back mode) but it can only respond to read requests from cache when the data page is present in the disk cache.
Contrary to the assumptions made in the previous work, HDDs do not have symmetric performance in all requests and their performance in random requests is asymmetric.
HDDs can serve random write requests with twofold performance compared to random read requests due to the internal caching mechanism.
HDDs have symmetric performance in sequential requests and the difference between read and write performance in strided requests is about 50\%.
The performance gap between strided and sequential requests is about 30\% in write requests and 200\% in read requests.
The reason for such performance improvement in sequential read requests is the read-ahead functionality in the disk controller.
This experiment shows that the assumptions made in previous work that HDDs have symmetric performance is partially valid but there are some cases that the performance difference between read and write requests is greater than 2x.

We have also examined the assumptions made in the previous work considering  that the request type does not affect the performance of SSDs and their performance in read requests is higher than in write requests.
Fig. \ref{fig:charperf} shows that although these assumptions are valid in most cases, there are numerous workloads in which these assumptions do not hold for SSDs.
For example, SSDs have higher performance in random write requests compared to random read requests, similar to HDDs.
The difference between these two, however, is less than their difference in HDDs which is about 60\%.
Apart from the mentioned difference, the performance of SSDs in read and write requests is similar to each other and the performance difference is less than 10\% in strided and sequential requests.
Therefore, strided requests should have higher priority for caching than sequential requests.
Although random write requests have less response time than random read requests in SSDs, the performance gain of caching random read requests is
higher than random write requests due to the performance difference of read and write requests in HDDs and SSDs.


\begin{figure}
	\centering
	\includegraphics[scale=0.79]{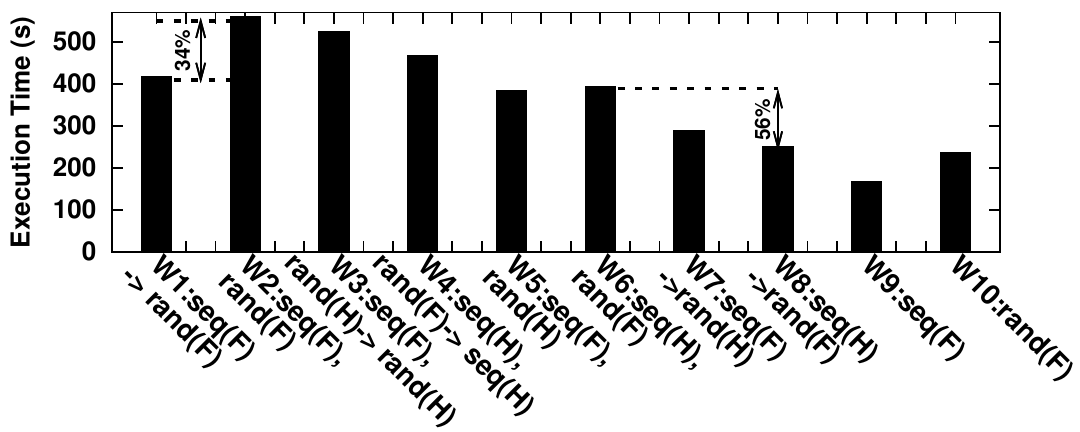}
	\vspace{-.2cm}
	\caption{Execution time of synthetic workloads on HDD}
	\label{fig:seqrand}
	\vspace{-.5cm}
\end{figure}

\begin{figure*}[!t]
\hspace{-0.3cm}
\subfloat[Rand. Cons.]{\includegraphics[width=.17\textwidth]{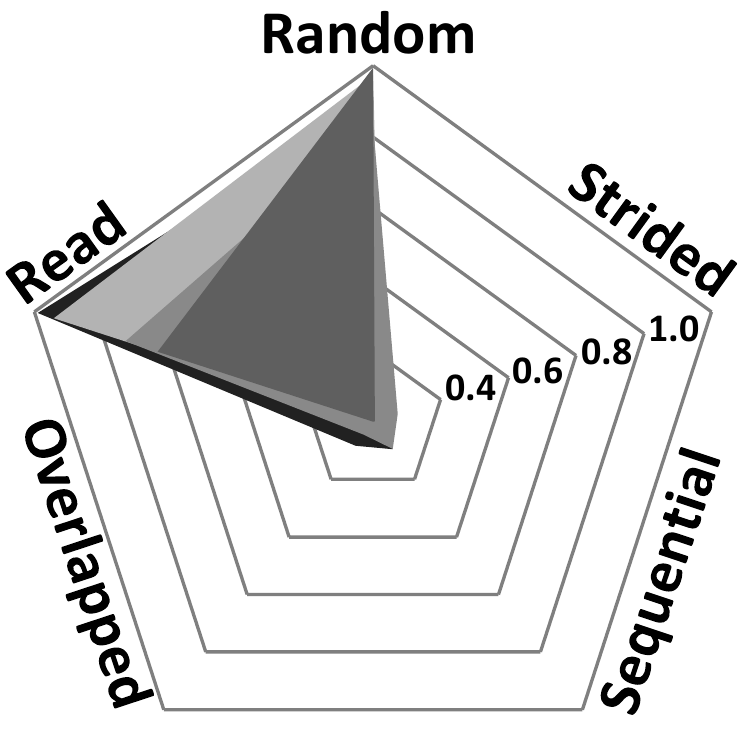}%
\label{fig:cat1}}
\subfloat[Seq. Prod./ Cons.]{\includegraphics[width=.17\textwidth]{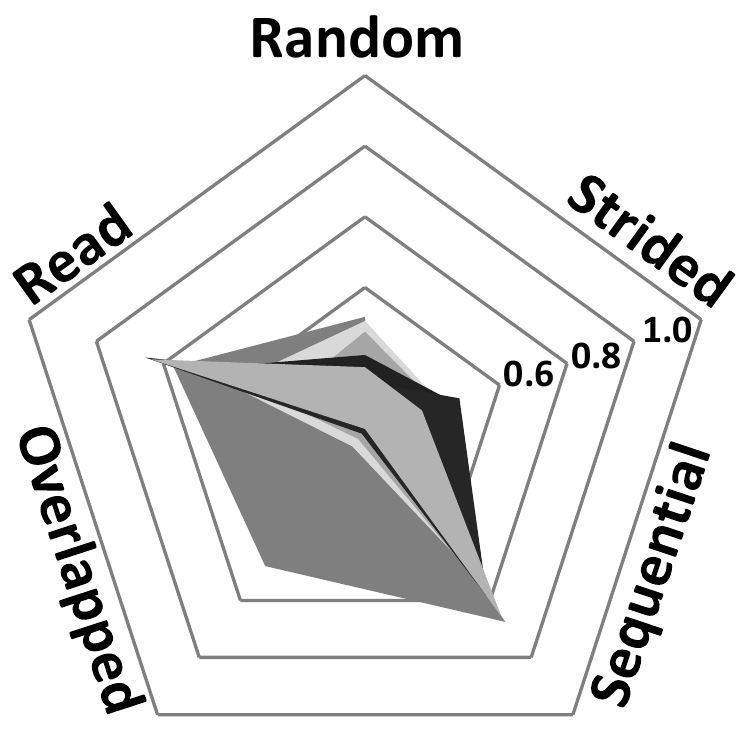}%
 \label{fig:cat2}}
\subfloat[Rand. Prod./ Cons.]{\includegraphics[width=.17\textwidth]{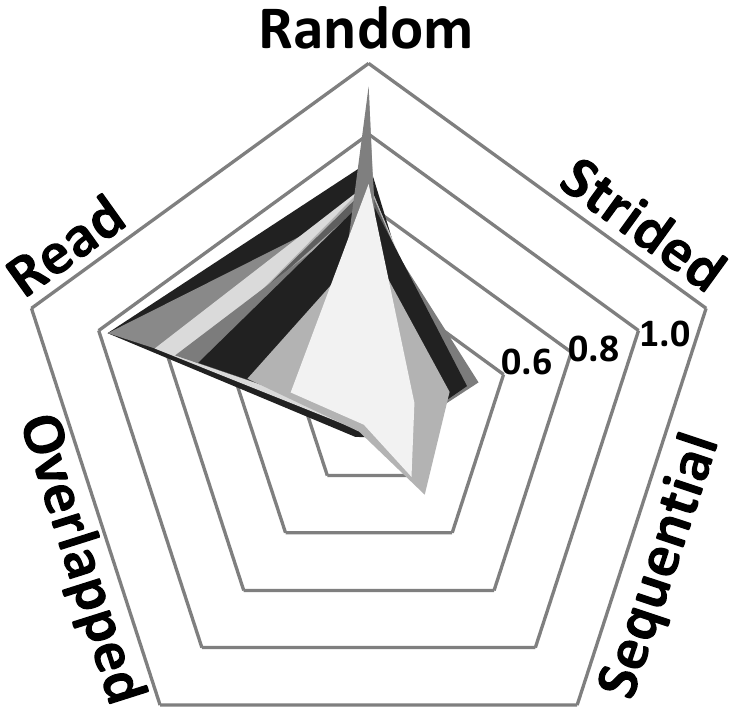}%
\label{fig:cat3}}
\subfloat[Archival Cons.]{\includegraphics[width=.17\textwidth]{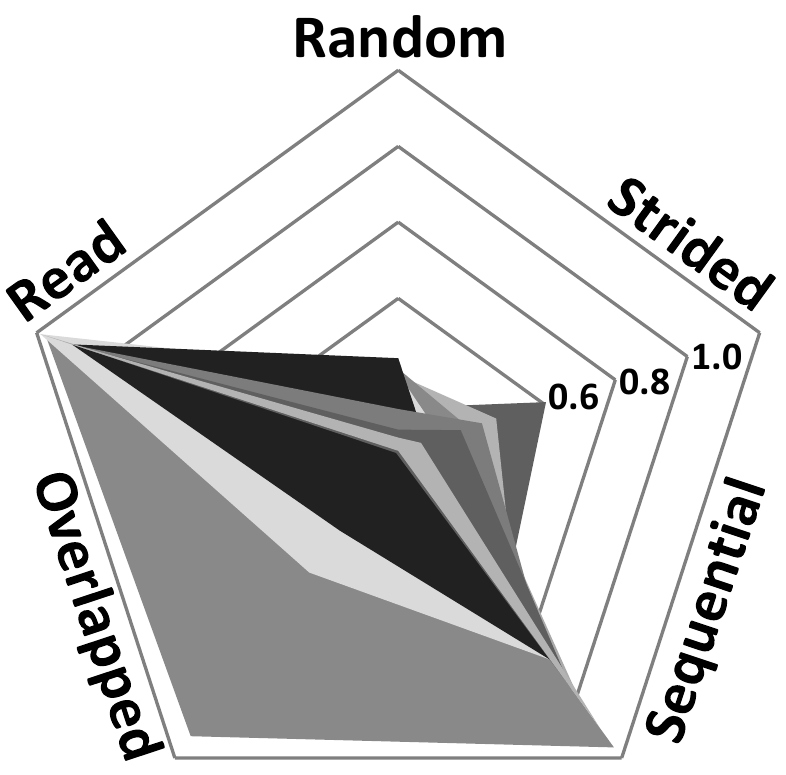}%
\label{fig:cat4}}
\subfloat[Large File Gen.]{\includegraphics[width=.17\textwidth]{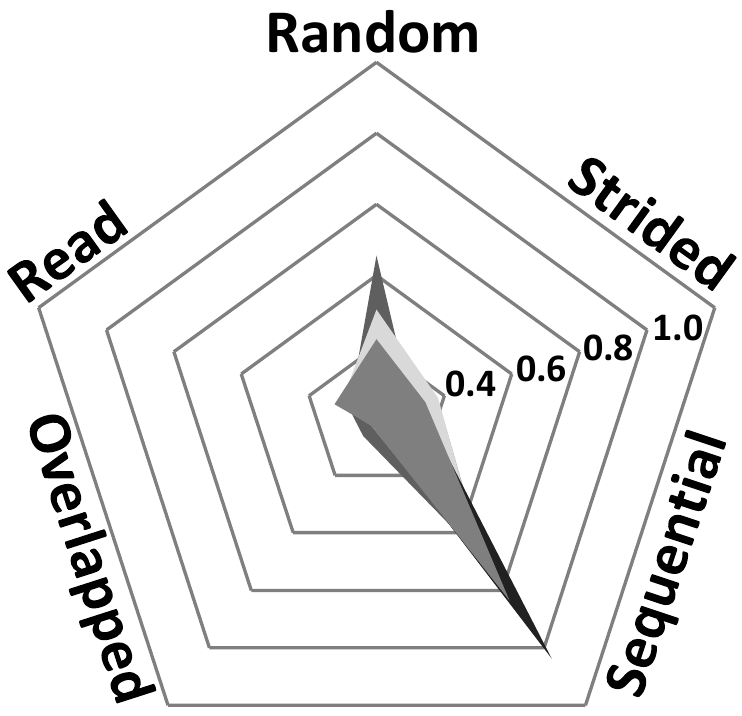}%
\label{fig:cat5}}
\subfloat[\small Overall]{\includegraphics[width=.17\textwidth]{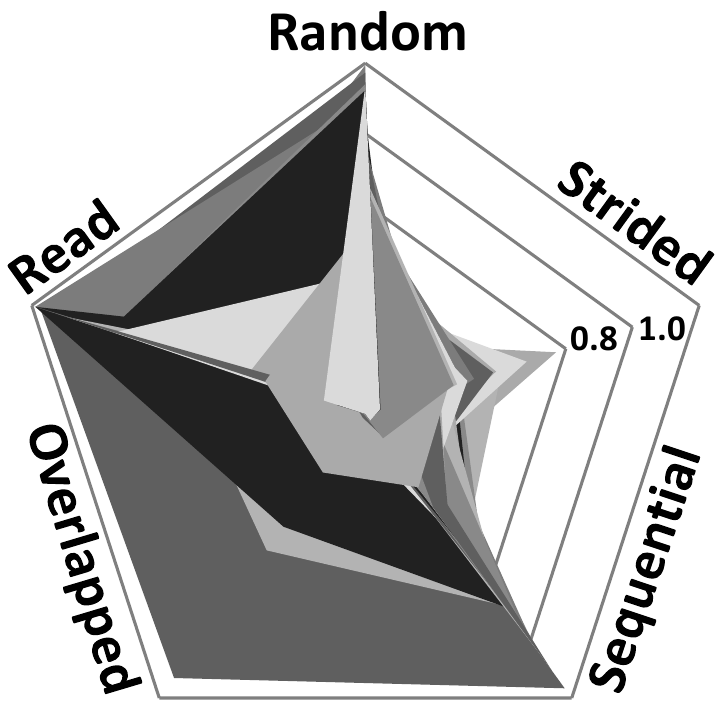}%
\label{fig:catall}}
\vspace{-0.3cm}
\caption{Workload characterization based on the request types}
\label{fig:char2}
\vspace{-.5cm}
\end{figure*}

\vspace{-.2cm}
\subsection{Request Interleaving}
\label{sec:interleaving}
Classifying requests is only a part of the workload characterization process and there are other important factors which should be considered in order to accurately predict the effect of each request on the performance.
The order of the requests and the interleaving of requests from different classes significantly affect the performance.
In order to show the effect of interleaved requests on performance, we have benchmarked HDDs with ten synthetic workloads consists of a set of random and sequential requests.
Each workload has two stages and the second stage will start as soon as all requests from the first stage are completed.
Each stage contains only either sequential, random, or both random and sequential requests.
Since the execution time of the workloads are similar to each other on SSDs, we only analyze the performance of HDDs for the sake of brevity.

Fig. \ref{fig:seqrand} shows the execution time of various synthetic workloads for HDD.
The ``-\textgreater'' sign shows the transition from the first to second stage and ``,'' denotes that the workloads are running simultaneously.
``F'' and ``H'' letters inside the parenthesis denote executing 20,000 and 10,000 requests, respectively.
Workloads W1 to W4 have the same number of requests, W5 and W7 have 50\% less random requests, and W6 and W8 have 50\% less sequential requests compared to the other workloads.
W2 has 34\% higher execution time compared to W1 although they have the same number of requests since the requests of W2 are interleaved with each other which results in more disk head movements.
The effect of interleaving of requests is related to the type of the interleaved requests and how much requests are interleaved, e.g., W3 and W4 have different execution time.
Another interesting observation in Fig. \ref{fig:seqrand} is that although W5 and W6 have different number of request types, their performance is similar since their requests are interleaved and the disk observes almost similar pattern for both.
The difference between the execution time of the interleaved requests depends on the ratio of the requests.
W1 and W2 have 34\% execution time difference while W6 and W8 have 56\% difference in their execution time.

\vspace{-.2cm}
\subsection{Workload Categories}
\label{sec:workloadchar}
In order to categorize I/O workloads \cite{snia1,tpccmanual,tpch,hammerDB,postmark,newfilebench}, we have explored many traces from various sources in addition to the traces obtained by running benchmark suits.
Over 75 traces have been analyzed and based on this analysis, five categories for workloads have been proposed which are a) random consumers, b) sequential producer/consumers, c) random producer/consumers, d) archival consumers, and e) large file generators.
In the remainder of this section, these categories will be detailed.

\vspace{-.2cm}
\subsubsection{Random Consumers:}
\emph{Random consumers} are read-dominant workloads that access the underlying storage device with random requests such as database management systems.
These workloads exhibit very low performance in HDD-based disk subsystem.
In addition, the write-back cache of HDDs which is able to cache random write requests and send them to the rotational magnetic part of the disk at a later time, is unable to perform well for \emph{random consumers} since most requests are read.
Fig. \ref{fig:cat1} shows the characterization of majority of analyzed workloads categorized as \emph{random consumers}.
The cache configuration for this category should have a small cache line size (same as application block size, e.g., 4-8KB) since almost all requests are random with a small size and, hence, using large cache line size will result in occupying the cache with undesirable data.
In addition, the access frequency is the most suitable metric for identifying performance-critical data pages since all request have the same size and request type.
Similarity of the requests means that there is no need for filtering a specified request type in order to cache or directly send them to the disk.

\vspace{-.2cm}
\subsubsection{Sequential Producer/Consumers:}
Most organizations have a sharing system based on network filesystems that users can access their files and share them with the others.
File types in this scenario are mostly documents that have a medium file size.
Since the underlying system is based on a filesystem, upon editing a file, the whole file contents will be uploaded to the server.
This will result in sequential accesses to the server which contains both read and write requests.
The optimal cache configuration for this category should consider the recency of accesses since users tend to access the newly created files more than old files.
The cache line size should preferably be set to larger values (128KB) since requests are mostly sequential and the eviction policy should be similar to LRU algorithm in order to catch the recency of accesses.
The degree of file sharing determines the number of overlapped requests in these workloads.
Fig. \ref{fig:cat2} depicts the characterization for this category which shows that the percentage of the overlapped requests varies inside this category.
The interleaving of sequential requests from accessing various files will result in a semi-random access pattern observed at the disk level since different files reside on different parts of the disk.
Therefore, identifying the I/O stream for each file and either caching all requests or ignoring them will help the cache to send more sequential requests to the disk and improve its performance.

\vspace{-.2cm}
\subsubsection{Random Producer/Consumers:}
Fig. \ref{fig:cat3} shows the characterization of the workloads in this category.
Although most requests in this category are random (similar to the \emph{random consumers} category), issuing many write requests by the applications in this category results in vast difference between two categories.
The performance gain on random read requests is almost twofold of random write requests as reported in Fig. \ref{fig:charperf}, which motivates us to favor random read requests more than random write requests for caching.
Our analysis shows that most mail servers and a few of \emph{On-Line Transaction Processing} (OLTP) servers belong to this category.
Therefore, in addition to the access frequency which is beneficial in identifying performance critical random requests, using recency is also helpful since for example, users access their new e-mails more often.
The cache configuration for this category consists of a small cache line size, a priority based eviction policy which assigns higher priority to read requests, and accumulates the priority based on access frequency.
This will enable us to capture recency, frequency, and read priority simultaneously.

\vspace{-.2cm}
\subsubsection{Archival Consumers:}
\emph{File Transfer Protocol} (FTP) and media servers have large number of concurrent users that access files for read purpose.
The size of the files in this category is quite large and almost all requests are sequential.
Fig. \ref{fig:cat4} depicts the percentage of various request types for \emph{archival consumers} category.
Since many concurrent users access the same set of files, the percentage of the overlapped requests is higher in this category compared to the \emph{sequential producer/consumers} category.
The concurrent users that access the same file may read different parts of the file at the same time which results in numerous strided requests.
The cache line size for this category can be set to a large value (128KB) in order to reduce the overhead of moving data pages to/from the cache.
The proposed eviction policy for this category is set to consider both frequency and recency of accesses to data pages.
The priority of each data page will be calculated based on its position in the LRU queue and the number of accesses to that data page.
Detecting sequential request streams can help to either cache or bypass cache for all requests of a stream.
This technique can reduce the randomness of the requests in both SSD and HDD which will improve performance.

\vspace{-.2cm}
\subsubsection{Large File Generators:}
Surveillance systems that store videos and monitoring systems have a few processes that write large files to the disk.
Almost all requests in the workloads in this category are writes and each process issues sequential requests.
Fig. \ref{fig:cat5} depicts the characterization of the workloads in this category.
Although all processes issue sequential requests, there exists many random requests in the examined workloads.
Our analysis shows that interleaving of the requests from various processes is the root cause of many sequential requests to appear as random in this category.
The proposed cache configuration for this category should have a large cache line size and should be configured as write-back mode.
The eviction policy for this category is similar to \emph{archival consumers} with a slight modification.
Upon evicting a data page, ReCA searches through the cache and all data pages with the physical address near the evicted data page will be evicted as well.
Since sequential requests have higher performance in HDDs, evicting data pages with physical address near each other will reduce the performance cost of cache flushing and ensures that the sequential requests from a process are committed at the same time to the HDD together and the disk observes less number of random requests.

\begin{figure*}[t!]
	\centering
	\includegraphics[scale=0.49]{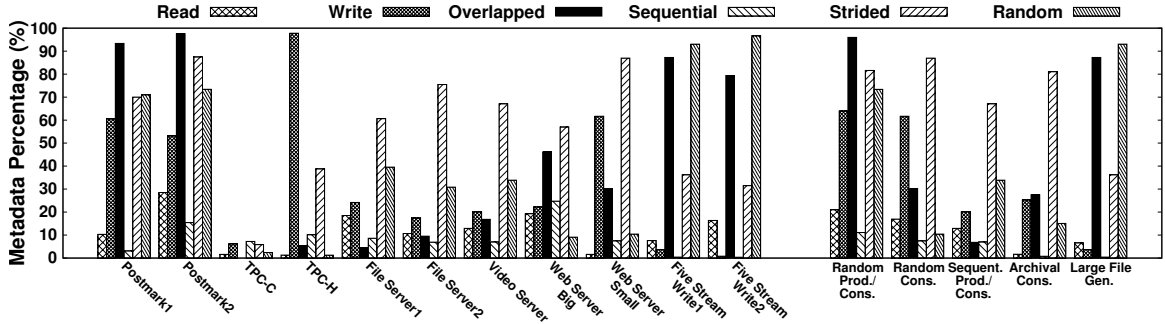}
	\vspace{-.3cm}
	\caption{Percentage of filesystem metadata requests in various request types of workloads}
	\label{fig:meta}
	\vspace{-.5cm}
\end{figure*}

\vspace{-.2cm}
\subsubsection{Workload Generality:}
Fig. \ref{fig:catall} shows the characterization for all analyzed workloads that cover majority of possible scenarios.
The only gap in the analyzed workloads is the lack of workloads with large number of strided requests.
This is due to the fact that applications are very unlikely to access the underlying disk with a strided pattern.
Database management systems, however, might access disk with strided pattern in order to read their indices.
The captured behavior of database management systems do not show this pattern since they try to optimize the requests issued to HDDs and prevent occurrence of such pattern.
ReCA is designed in such a way that new workload categories and their optimal cache configuration can be added even in runtime without any architecture modifications.
Therefore, if new applications with unusual behavior are emerged, they can be added to ReCA to maintain its workload generality.

\begin{figure}[t]
	\hspace{-0.8cm}
	\includegraphics[scale=0.33]{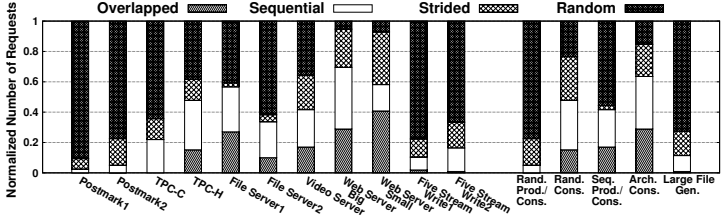}
	\vspace{-.5cm}
	\caption{Filesystem metadata request type breakdown}
	\label{fig:metadatabreakdown}
	\vspace{-.5cm}
\end{figure}

\vspace{-.2cm}
\subsection{Filesystem Metadata}
\label{sec:metadata}
Filesystem issues many requests for managing its internal data structures and storing additional information for each file.
These additional data blocks are called filesystem metadata blocks.
Since each metadata block contains information about numerous number of user data blocks and in addition prior to accessing a user data block, its corresponding metadata block should have been accessed, the metadata requests have higher impact on the overall performance than individual user requests.
The importance of filesystem metadata from performance perspective has been analyzed and demonstrated in previous studies \cite{Hystor, Azor, tiering}.
These studies, however, neglect further analysis of filesystem metadata, its various types, and the characterization of filesystem metadata in different workload types.
For example, in reading sequential data blocks from a file, only one metadata request (to read an indirect block) is required to find physical location of 1024 data blocks.
Reading random data blocks from a file, however, requires reading one indirect block per data block since requests are scattered across file logical addresses.

Fig. \ref{fig:metadatabreakdown} depicts the characterization of the filesystem metadata for many representative workloads and the average of all workloads for each workload category.
The overlapped requests have been considered with the other request types in Fig. \ref{fig:metadatabreakdown} for simplicity.
If a request is flagged as overlapped, its main type is ignored in order to present all request types in one figure.
As shown in Fig. \ref{fig:metadatabreakdown}, different workload categories have different metadata access patterns which is in contrary to the assumptions made in the previous work \cite{tiering,Hystor}.
Random consumers category, despite of being a random workload, has low percentage of random filesystem metadata requests.
Our careful analysis shows that applications in the \emph{random consumers} category have less locality than applications in \emph{random producer/consumers}.
Higher locality results in issuing less metadata requests to disks since filesystem tries to cache metadata information and each cached metadata block can serve more requests if the workload demonstrates higher locality.
Therefore, in workloads with higher locality, the requested metadata blocks from disk (which are less in number) will be more random.
In workloads with less locality, more metadata requests will be issued to disk and since the filesystem tries to place metadata requests close to the actual user data,
requests will be less random and instead more sequential or strided.
Such important observation will clarify the difference between \emph{random consumers} and \emph{random producer/consumers}.
Workload categories have other dissimilarities such as different strided requests percentage as well.
The mentioned differences support our claim about the need for treating filesystem metadata requests differently across workload categories.

To further analyze the access pattern of the filesystem metadata requests, the percentage of the filesystem metadata requests for each request type is depicted in Fig. \ref{fig:meta}.
The contribution of the filesystem metadata differs vastly across workload categories which makes using one single rule for prioritizing them not accurate enough.
For example, both of the \emph{large file generators} and \emph{archival consumers} categories are sequential workloads, however, the breakdown of the filesystem metadata requests differs vastly across them.
One common pattern among all workload categories is that most strided requests are for filesystem metadata which confirms our assumption that applications do not issue many strided requests to disk.
Such information enabled us to further optimize the workload characterization process and prioritize filesystem metadata requests based on their impact on the overall performance for each workload category.

It is noteworthy that identifying filesystem metadata at the block I/O layer in the OS is not possible since filesystems do not leverage these information.
To this end, previous studies tried to modify the kernel structures in order to pass the mentioned information to the lower layers of I/O stack.
This approach is possible in the proposed workload characterization.
If modifying OS layers is intractable, previous work will be unlikely to consider the filesystem metadata.
The proposed workload characterization, however, can use the average probability that a request is filesystem metadata based on the information provided in Fig. \ref{fig:meta}.
This makes the proposed workload characterization more accurate than the previous studies in identifying filesystem metadata requests.

\begin{figure}
\centering
\includegraphics[scale=0.675]{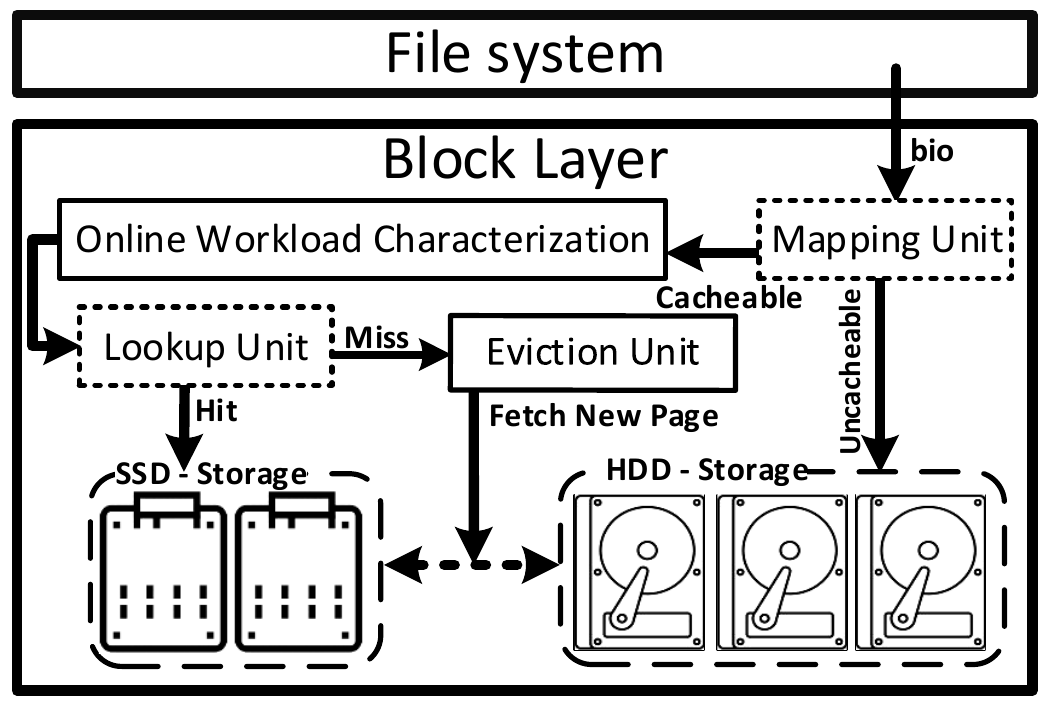}
\vspace{-.2cm}
\caption{The proposed storage architecture (ReCA)}
\label{fig:rfc1}
\vspace{-.5cm}
\end{figure}

\vspace{-.3cm}
\section{Proposed Architecture}
\label{sec:proposed}
In the proposed architecture, we monitor the running workload and based on the observed request types, the most suitable workload category is selected for the running workload.
We then compare the running workload with the predefined workload categories based on the percentage of the request types and the interleaving of the requests.
Based on the selected category, the cache configuration will be set to one of the predefined configurations presented earlier in Section \ref{sec:workloadchar}.
The cache reconfiguration will be done online and the users will not notice the reconfiguration process, except for the performance improvement upon completing the reconfiguration.

\vspace{-.2cm}
\subsection{Overall architecture}
Fig. \ref{fig:rfc1} shows the high level architecture of ReCA which consists of four major modules: a) \emph{mapping unit}, b) \emph{online workload characterization unit}, c) \emph{lookup unit}, and d) \emph{eviction unit}.
Two units with dashed-line boxes are general units which exist in almost all caching architectures while units with solid lines boxes are presented in the proposed architecture.
The mapping unit receives requests from upper layers and breaks them into smaller requests with the same size as the cache line size.
If the request is tagged as \emph{uncachable}, the mapping unit will send it directly to the disk and if the corresponding data page exists in the cache, it invalidates its block.
The lookup unit searches through the cache and if it finds the requested data page, it redirects the request to SSDs and otherwise, it is sent to HDDs.
\emph{Online workload characterization unit} monitors the arrived requests and tries to find the best suited  workload category and based on these information, it updates the internal data structures and reconfigures the cache.
Contrary to the conventional caching algorithms that move a data page to the cache in case of a miss, ReCA either does not move the page to the cache or will move it at a later time which is decided by the eviction unit in the proposed architecture.
The eviction unit receives the miss requests and decides whether it should be placed in cache or not.
This unit also decides which data pages are no longer needed in the cache and can be evicted in order to free up cache space for new incoming data pages.

\begin{figure}
\hspace{-0.35cm}
\includegraphics[scale=0.51]{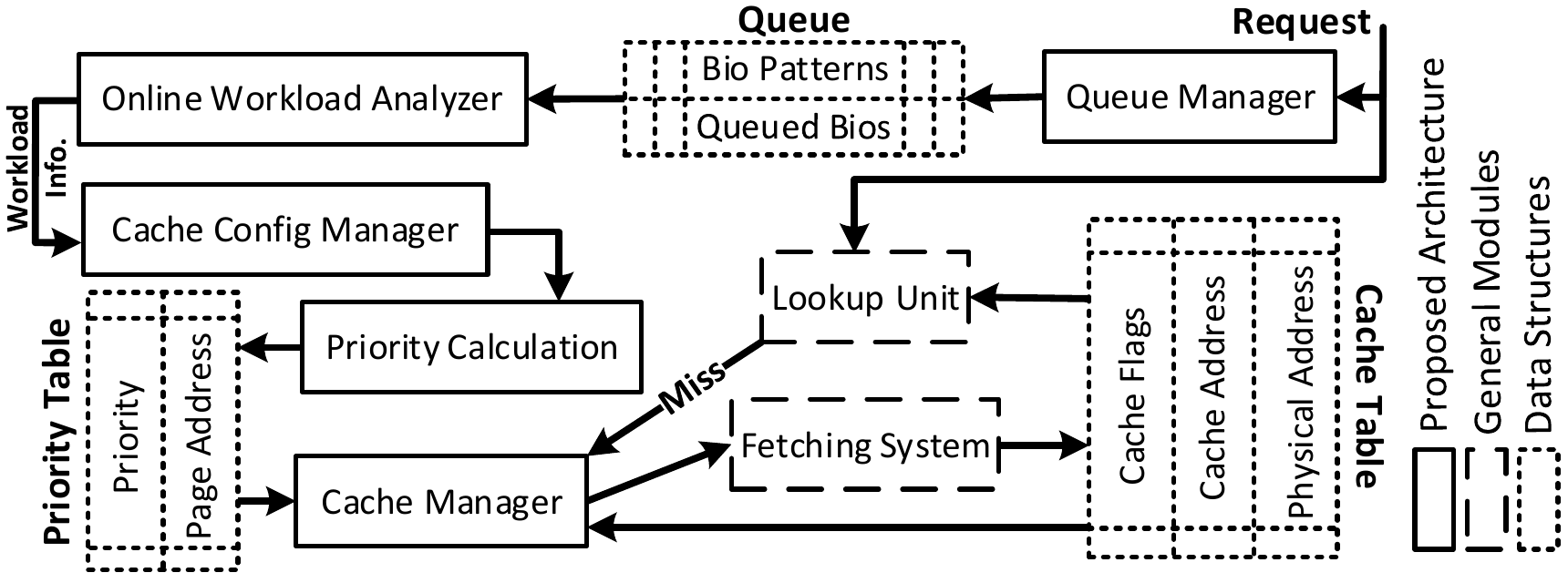}
\vspace{-0.2cm}
\caption{ReCA: Internal details}
\label{fig:rfc2}
\vspace{-.20cm}
\end{figure}

Fig. \ref{fig:rfc2} depicts the detailed modules and data structures used in ReCA.
The boxes with dotted lines show the data structures, boxes with dashed lines are for general modules, and boxes with solid lines depict the modules designed for the proposed architecture.
In order to identify the request type, the queue manager puts the requests into a queue with length of 64 to compare them with the previous requests and decide their types.
\emph{Lookup} unit looks through the \emph{cache table} in order to find the requested data pages.
If a data page is not available in the cache, the \emph{cache manager} will be notified.
This module fetches the priority of the requested data page and if it has higher priority than a data page which resides in the cache, it calls the \emph{fetching system} to replace the data page in cache and updates the \emph{cache table}.
\emph{Online workload analyzer} monitors the incoming requests and if the pattern of the requests changes, the \emph{cache configuration manager} will be notified to change the \emph{priority table} to reflect the change in the workload.
The monitoring function can detect phases within an application.
In addition, multi process applications working with the same data will be recognized by the monitoring function.
\emph{Priority table} is managed using \emph{B-tree} and can be expanded up to 50\% of the total free memory.
\emph{Priority calculation} unit also will be notified to change the algorithm for calculating the priority.
Since each workload category has a unique algorithm for deciding the performance criticality of each data page, the history table includes five columns for each data page which imposes significant overhead to the system.
In order to reduce this overhead, at any given time, only one column of the history table is active and when the workload category changes, the corresponding data structures will be freed.

\newlength{\oldtextfloatsep}\setlength{\oldtextfloatsep}{\textfloatsep}
\setlength{\textfloatsep}{0pt}
\begin{algorithm}[t]
\caption{Caching Algorithm}\label{alg:caching}
\scriptsize
\begin{algorithmic}[1]
\Procedure{Lookup Function}{} \label{alg:lookupstart}
\State Upon reaching of a new I/O as \textit{new\_io}
\State $Access\_History\_Manager(\textit{new\_io})$ \label{alg:callaccesshist}
\For {each \textbf{cache\_line\_size} block accessed through \textit{new\_io} as \textit{blk}}
\If {\textit{blk} is in \textit{cache\_table}}
\State $\textit{cache\_table.Priority}[\textit{blk.addr}] \gets \textit{history\_table.Priority}[\textit{blk.addr}]$
\State issue request to cache
\Else
\State $\textit{cache\_blk} \gets Cache\_Manager(\textit{blk})$
\If {\textit{cache\_blk} != \textit{NULL}}
\State $\textit{cache\_table}[\textit{cache\_blk}] \gets HDD[\textit{blk.addr}]$
\EndIf \State\textbf{end if}
\EndIf \State\textbf{end if}
\EndFor \State\textbf{end for}
\EndProcedure \label{alg:lookupend}
\Procedure{Cache\_Manager(\textit{blk})}{} 
\For {every block in \textit{cache\_table} as \textit{blk}}
\If {\textit{blk}.Flag == \textit{INVALID}}
\State \textbf{Return} \textit{blk}
\EndIf \State\textbf{end if}
\EndFor \State\textbf{end for}
\State $\textit{min\_pr\_blk} \gets$ page cache block with minimum priority
\If {\textit{cache\_priority\_table}[\textit{min\_pr\_blk}] $<$ \textit{priority\_table}[\textit{blk.addr}]}
\If {\textit{cache\_table}[\textit{min\_pr\_blk}].flag == \textit{DIRTY}}
\State Flush \textit{cache\_table}[\textit{min\_pr\_blk}] to HDD
\EndIf \State\textbf{end if}
\State \textbf{Return} \textit{min\_pr\_blk}
\Else
\State \textbf{Return} \textit{NULL}
\EndIf \State\textbf{end if}
\EndProcedure
\Procedure{Access\_History\_Manager(\textit{new\_io})}{}
\State $\textit{Queue}.\text{Push}(new\_io)$
\If {$\textit{Queue.length} > Queue\_Length\_Threshold$} 
\State $\textit{old\_io} \gets \text{Pop}(Queue)$
\State $\textit{io\_type} \gets Characterization(old\_io)$
\For {each \textbf{cache\_line\_size} block accessed through \textit{old\_io} as \textit{blk}}
\State \textit{priority\_table}[\textit{blk.addr}] += Priority\_Calc(\textit{io\_type})
\EndFor \State\textbf{end for} 
\EndIf \State\textbf{end if}
\State \textit{io\_list}.append(\textit{io\_type}) \label{alg:callcharstart}
\If {$\textit{io\_list}.length > Workload\_Check\_Threshold$}
\State $\textit{Analyze\_Workload}(\textit{io\_list})$ 
\EndIf \State\textbf{end if} \label{alg:callcharend}
\EndProcedure
\end{algorithmic}
\end{algorithm}

\vspace{-.2cm}
\subsection{Proposed Algorithms}
ReCA uses three main procedures to manage the cache internals and three other procedures for characterization of the workloads and reconfiguring the cache.
Algorithm \ref{alg:caching} shows the procedures for managing the cache, i.e., \emph{lookup\_function}, \emph{cache\_manager}, and \emph{access\_history\_manager}.
The former procedure outlined in lines \ref{alg:lookupstart} through \ref{alg:lookupend}, is called for each incoming request.
If the request size is greater than the cache line size, it will be divided into several subrequests.
The incoming request will be responded after all subrequests have been served.
If an error occurs during processing of a subrequest, the request will be replied by an error code (not shown in Algorithm \ref{alg:caching}).
\emph{Lookup function} searches through the cache and tries to serve as many as possible requests from cache.
Since this function is the main entry point of ReCA, \emph{Access History Manager} is called from this function (line \ref{alg:callaccesshist}) in order to update the request queue.
When a miss occurs, \emph{Cache Manager} will be called in order to decide whether or not the requested data page should be moved to SSD.
If the cache has empty space, the page will be moved to SSD for future references.
When the cache is fully occupied, the priority of the requested data page will be compared to the priority of the data pages residing in the cache and if the requested data page has higher priority, it will replace the data page with minimum priority.
Since the overhead of managing and comparing the priorities becomes significant in case of large cache size, ReCA divides the cache into several sets and only compares the priorities within a set.
Having a low hit ratio in the cache can result in high number of pending jobs for moving data pages between disk and SSD which will impose significant overhead.
In order to limit this overhead, the number of pending jobs is limited and queued jobs will be started asynchronously.
\emph{Access History Manager} manages request queue which is used for determining the workload category.
Upon evicting a data page from the queue, the priority of the data page will be updated.
In lines \ref{alg:callcharstart} through \ref{alg:callcharend}, \emph{Analyze\_Workload} is called periodically after processing a number of requests in order to analyze the current workload characteristics and change the workload type if necessary.
The number of processed requests between calls to \emph{Analyze\_Workload} determines the reaction time of ReCA to the changes in the workload.
Small values trigger reconfiguration more frequently which incur more performance overhead to the system while using large values delays identifying the change in the workload.
The value used for \emph{Workload\_Check\_Threshold} in ReCA is set to 100,000 requests.

The main characterization and reconfiguration processes used in ReCA are shows in Algorithm \ref{alg:characterization}.
\emph{Characterization} function analyzes a request based on the other requests in a queue and identifies its type.
Characterizing individual requests is the first step of the workload characterization process which is outlined in lines  \ref{alg:charstart} through \ref{alg:charend} based on the request classification presented in Section \ref{sec:reqclass}.
The next step after characterizing the requests is prioritizing them based on their performance on storage devices which is done in \emph{Priority\_calc}.
This function calculates the assigned priority for a request and adds it to the priority of the corresponding data page.
The calculated priority consists of overlapped (if request is flagged as overlapped), access type, and read/write priorities.
In the implementation, read/write priority is set to be dependent to access type in order to have more flexibility in assigning priorities (not shown in Algorithm \ref{alg:characterization}).
The priority values are extracted from predefined priorities of the workload categories which are based on the analysis discussed in Section \ref{sec:workloadchar}.
New workload categories with their corresponding priorities and optimal cache configurations can be added to \emph{characteristics table} in runtime to extend the characterization capabilities of ReCA.

The actual reconfiguration is done in \emph{Analyze\_workload} function which first decides the current workload type and then triggers the reconfiguration process if necessary.
The proposed cache reconfiguration process is online without disturbing application I/Os.
The cache will not be flushed entirely and only the required data pages are brought to the cache and  lower priority data pages are evicted.
Workload identification is based on the predefined workload types and the collected data from the current workload.
ReCA calculates the \emph{Euclidean} distance between the current workload type and each of the predefined workload types.
The workload type with minimum Euclidean distance will be selected as the new current workload.
After choosing the workload type, cache data structures will be reconfigured.
The reconfiguration process starts by changing the priorities of data pages based on the new workload type.
After updating the priorities, many data pages will have lower priorities which results in eviction from cache.
The limitation employed in the number of concurrent evictions prevents the performance degradations because of the large number of evictions after updating the workload type.
Hence, the performance impact of cache reconfiguration is limited and applications will not observe any performance degradation.
ReCA can have different eviction policies based on the workload type which is set in Line \ref{alg:evictionpolicyset} of Algorithm \ref{alg:characterization}.
Workload types 
can have different cache line sizes which are optimized towards their characteristics.

In case of running multiple applications simultaneously, ReCA can identify category of each application, separately by tagging the process  \emph{Identification Number} (ID) of applications in \emph{(io list)} and calculating the workload type for each process.
This technique can also be employed in virtualization environments by characterizing each virtual machine category.
The memory overhead of maintaining required data structures is negligible and does not affect the performance of ReCA.

\begin{algorithm}[t]
\caption{Characterization Algorithm}\label{alg:characterization}
\scriptsize
\begin{algorithmic}[1]
\Procedure{Characterization(old\_io)}{}  \label{alg:charstart}
\For {each I/O in \textit{Queue} as \textit{temp\_io}}
\If {\textit{old\_io}.size $>$ $Seq\_threshold$ or \textit{old\_io}.end == \textit{temp\_io}.begin}
\State $\textit{result.access} \gets \textbf{sequential}$
\State $\textbf{Break}$
\ElsIf{(\textit{old\_io}.begin $>$ \textit{temp\_io}.end and \textit{old\_io}.begin - \textit{temp\_io}.end $<$ $Strd\_threshold$) or (\textit{old\_io}.end $<$ \textit{temp\_io}.begin and \textit{temp\_io}.begin - \textit{old\_io}.end $<$ $Strd\_threshold$)}
\State $\textit{result.access} \gets \textbf{strided}$
\EndIf \State\textbf{end if}
\If{\textit{(old\_io}.begin $>$ \textit{temp\_io}.begin and \textit{old\_io}.begin $<$ \textit{temp\_io}.end) or (\textit{old\_io}.end $>$ \textit{temp\_io}.begin and \textit{old\_io}.end $<$ \textit{temp\_io}.end)}
\State $\textit{result.isOver} \gets \textbf{True}$
\EndIf \State\textbf{end if}
\EndFor \State\textbf{end for}
\If {\textit{result.access} == \textit{null}}
\State $\textit{result.access} \gets \textbf{random}$
\EndIf \State\textbf{end if}
\If {\textit{old\_io}.type equals \textit{write}}
\State $\textit{result.R\_Wr} \gets \textbf{WRITE}$
\Else
\State $\textit{result.R\_Wr} \gets \textbf{READ}$
\EndIf \State\textbf{end if}
\State \textbf{Return} $\textit{result}$
\EndProcedure  \label{alg:charend}
\Procedure{Priority\_Calc(\textit{io\_type})}{}
\State $result \gets 0$
\If {\textit{io\_type.isOver}}
\State \textit{result} += \textit{characteristics\_table[current\_workload]}.over\_priority
\EndIf \State\textbf{end if}
\State \textit{result} +=  \textit{characteristics\_table[current\_workload]}.acc\_priority[\textit{io\_type.access}]
\State \textit{result} +=  \textit{characteristics\_table[current\_workload]}.rw\_priority[\textit{io\_type.io\_type.RW}]
\State \textbf{Return} $\textit{result}$
\EndProcedure
\Procedure{Analyze\_Workload(\textit{io\_list})}{}
\State $current\_workload$ $\gets$ workload type with minimum Euclidean distance to \textit{io\_list}
\State Update $Priority\_Table$ based on $current\_workload$
\State  $eviction\_policy  \gets  $ \textit{characteristics\_table[current\_workload]}.eviction\_policy \label{alg:evictionpolicyset}
\State Reconstruct $cache\_table$ for new $cache\_line\_size$
\EndProcedure
\end{algorithmic}
\end{algorithm}

When $current\_workload$ is updated due to changes in the workload characteristics, the cache line size may need to be updated as well.
Since both $cache\_table$ and $priority\_table$ are dependent to the cache line size, ReCA reconstructs these tables.
In case of updating from a larger cache line size to a smaller cache line size, ReCA extends each entry into many entries and sets the priority of new entries to $\frac{original\_priority}{\#\_of\_new\_entries}$.
Therefore, the total value of priorities in the $priority\_table$ remains the same.
Migrating from a small cache line size to a larger cache line size, however, is not trivial since a large cache line size spans across many data pages and if one of the data pages belonging to a cache line does not exist in the cache, the incoming request will not be responded. 
To fix this issue, ReCA fills the missing data pages asynchronously and stores a bit for each data page in cache lines in order to identify the missing data pages.
If a request for a missing data page arrives, ReCA (a) issues a synchronous request to the disk subsystem, (b) retrieves the data page, and then (c) responds the user request.
The retrieved data page will be sent to SSD in order to fill the missing data page.

\vspace{-.3cm}
\section{Workflow}
\label{sec:workflow}
\vspace{-.05cm}
In this section, the overall workflow of ReCA is presented.
The workflow which enables ReCA to be reconfigurable in the runtime with optimized configuration regarding the current workload consists of a set of offline analysis and a comprehensive characterization in addition to a novel online process for cache optimization.
The overall workflow of ReCA consists of offline and online processes is demonstrated in Fig. \ref{fig:workflow}.
The goal of offline processes is to investigate the performance of storage devices over real-world applications and cache configurations and then decide the optimal cache configuration for each application type.
The online processes try to \emph{(a)} monitor the currently running application, \emph{(b)} identify its type, and \emph{(c)} reconfigure cache architecture to best suit the application performance requirements.

\vspace{-.35cm}
\subsection{Offline Processes in ReCA}
\begin{figure}
	\centering
	\includegraphics[scale=0.515]{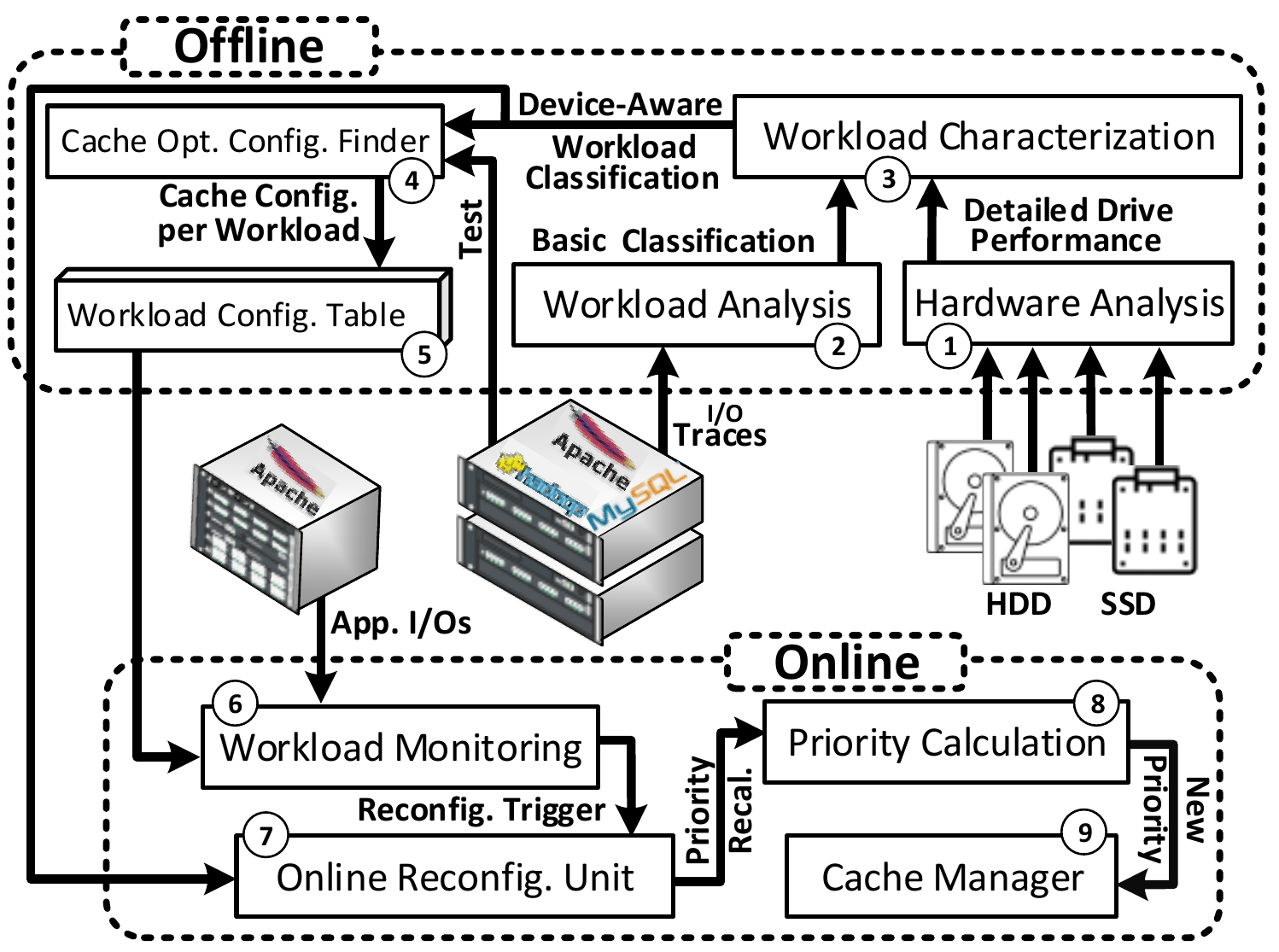}
	\vspace{-.2cm}
	\caption{ReCA workflow}
	\label{fig:workflow}
	\vspace{-.53cm}
\end{figure}

To determine an optimal cache configuration for various applications, the storage devices are extensively examined using synthetic and real-world applications (\emph{Hardware Analysis} in Fig. \ref{fig:workflow}).
A wide range of enterprise applications are analyzed and categorized in the next step to investigate their I/O behavior (\emph{Workload Analysis} in Fig. \ref{fig:workflow}).
Since many of the inner-workload interactions between requests cannot be evaluated without actual experiments, another step is added to the offline processes (\emph{Workloads Characterization} in Fig. \ref{fig:workflow}) in which the actual performance of various applications is tested over real storage devices to determine their actual performance.
The output of \emph{Workloads Characterization} determines the actual workload categories used in ReCA along their performance characteristics.
In \emph{Cache Opt. Config. Finder} unit, enterprise applications will be executed under various cache architectures and policies in order to determine the optimal cache configuration for the target application type.
Such cache architectures and policies are given to the ReCA online section by filling \emph{Workload Config. Table}.

\vspace{-.3cm}
\subsection{Online Processes in ReCA}
In addition to the traditional modules of a caching architecture, ReCA employs other modules to detect the application type and reconfigure the cache accordingly.
Currently running application is monitored in \emph{Workload Monitoring} and upon detection of any change in the workload behavior (based on predefined workload types in \emph{Workload Config. Table}), the reconfiguration process is triggered.
The reconfiguration process takes the current and the target cache configurations into account and plans the reconfiguration with minimal performance overhead and no downtime for application I/Os.
The internal priority and history of data pages is also updated to reflect the changes in the application type.

\setlength{\textfloatsep}{\oldtextfloatsep}

\begin{figure*}[!t]
	\centering
	\includegraphics[width=0.75\textwidth]{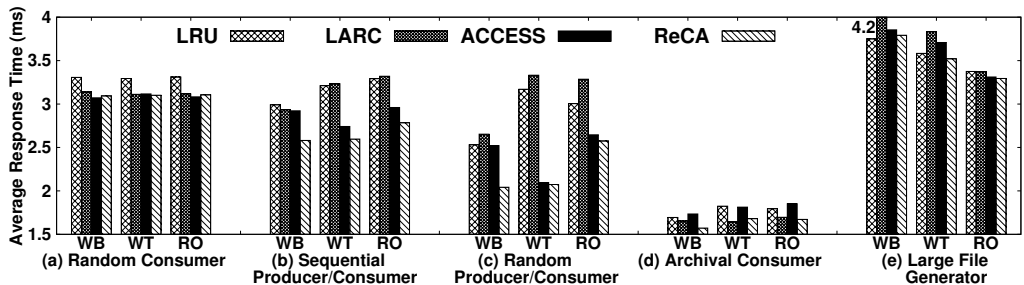}%
	\vspace{-.23cm}
	\caption{Average Response Time in various Write Policies}
	\label{fig:wp}
	\vspace{-.53cm}
\end{figure*}

\vspace{-.3cm}
\section{Experimental Results}
\label{sec:results}
\vspace{-.1cm}
This section presents the experimental setup and results for evaluating ReCA against previous caching architectures.
Section \ref{sec:experimentalsetup} describes the experimental setup.
The evaluations are presented in Section \ref{sec:experimentalresults}.
Parameter sensitivity is discussed in Section \ref{sec:sensitivity}.
Section \ref{sec:mixedworkload} reports ReCA performance under multiple running applications.
Performance overhead is presented in Section \ref{sec:perfoverheads}.
Finally, the performance of running ReCA in tiering mode is discussed in Section \ref{sec:tieringreca}.

\vspace{-.3cm}
\subsection{Experimental Setup}
\label{sec:experimentalsetup}
\vspace{-.1cm}
In order to accurately evaluate ReCA, all experiments in this paper have been performed in a physical server with Xeon E5620 CPU and 32 GB of main memory.
The operating system of the testbed was Ubuntu 14.04 running Linux kernel 3.17.0.
The specification of the employed HDDs and SSDs is presented in Table \ref{tbl:config}.
The code base for implementation of ReCA\footnote{The source code of ReCA is publicly available with the same license as EnhanceIO at \url{http://dsn.ce.sharif.edu}.} is EnhanceIO which is an open source caching solution \cite{enhanceio}.
In all experiments, cache size is set to 20\% of the total unique data pages in the workload, unless explicitly said otherwise.
All architectures were given enough cache warm-up time to reach a stable state, roughly 10-20 minutes. 
ReCA has been compared with LARC algorithm which is a variation of LRU algorithm proposed specifically for SSD caching as opposed to many of the other LRU variations that have general purpose.
This makes LARC best suited for comparing with ReCA.
In addition to LARC, the proposed architecture is compared to an access frequency algorithm that counts the number of accesses to data pages.
This algorithm represents the frequency algorithms while LARC represents the recency algorithms.
Note that the access frequency algorithm counts the number of accesses to all data pages in the target address space which makes its memory overhead significantly higher than ReCA and LARC.
To further show the efficiency of each algorithm, a baseline LRU algorithm is also considered in the experiments.
Since the performance of SSDs varies as they age, SSDs are trimmed after each experiment using \emph{blkdiscard} utility in Linux operating system.
Due to the effect of request scheduler and garbage collection in SSDs, response time of requests might have considerable variations.
	To reduce the performance variations, experiments had long execution time so that SSD reaches a steady state with small variation in requests response time.
	By repeating the experiments, we have made sure that the variations have negligible effect on the results.

\vspace{-.3cm}
\subsection{Experimental Results}
\label{sec:experimentalresults}
\vspace{-.1cm}
Fig. \ref{fig:wp} shows the average response time of caching architectures under various write policies and workload categories.
ReCA outperforms LARC algorithm in almost all workload categories and write policies which approves the efficiency of the proposed algorithm for prioritizing data pages.
The difference between performance of various write policies is not the same across workload categories.
In \emph{Random Consumers} category, the difference is negligible while \emph{Large File Generators} category has significant performance gap between different write policies.
Combination of this observation and the fact that write policies have different endurance and reliability costs (read-only extends cache lifetime while write-back shortens SSD lifetime but increases data loss probability), supports our claim that cache write policy should be adaptive.

\begin{figure}[!t]
\hspace{-1.6cm}
\centering
\includegraphics[width=0.54\textwidth]{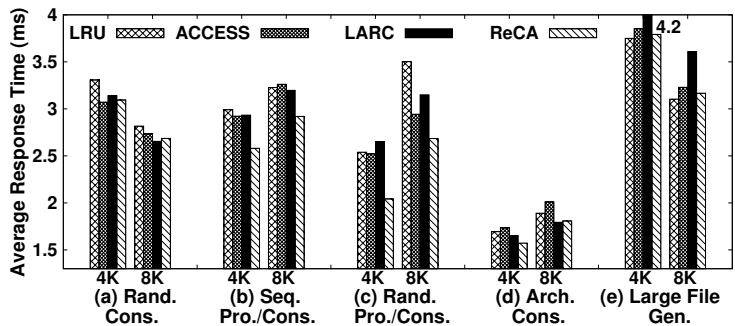}%
\vspace{-.22cm}
\caption{Average Response Time of Various Cache Architectures for two Cache Line Sizes (4KB, 8KB)}
\label{fig:cl}
\vspace{-.5cm}
\end{figure}

\begin{figure}[t]
\centering
\includegraphics[scale=0.53]{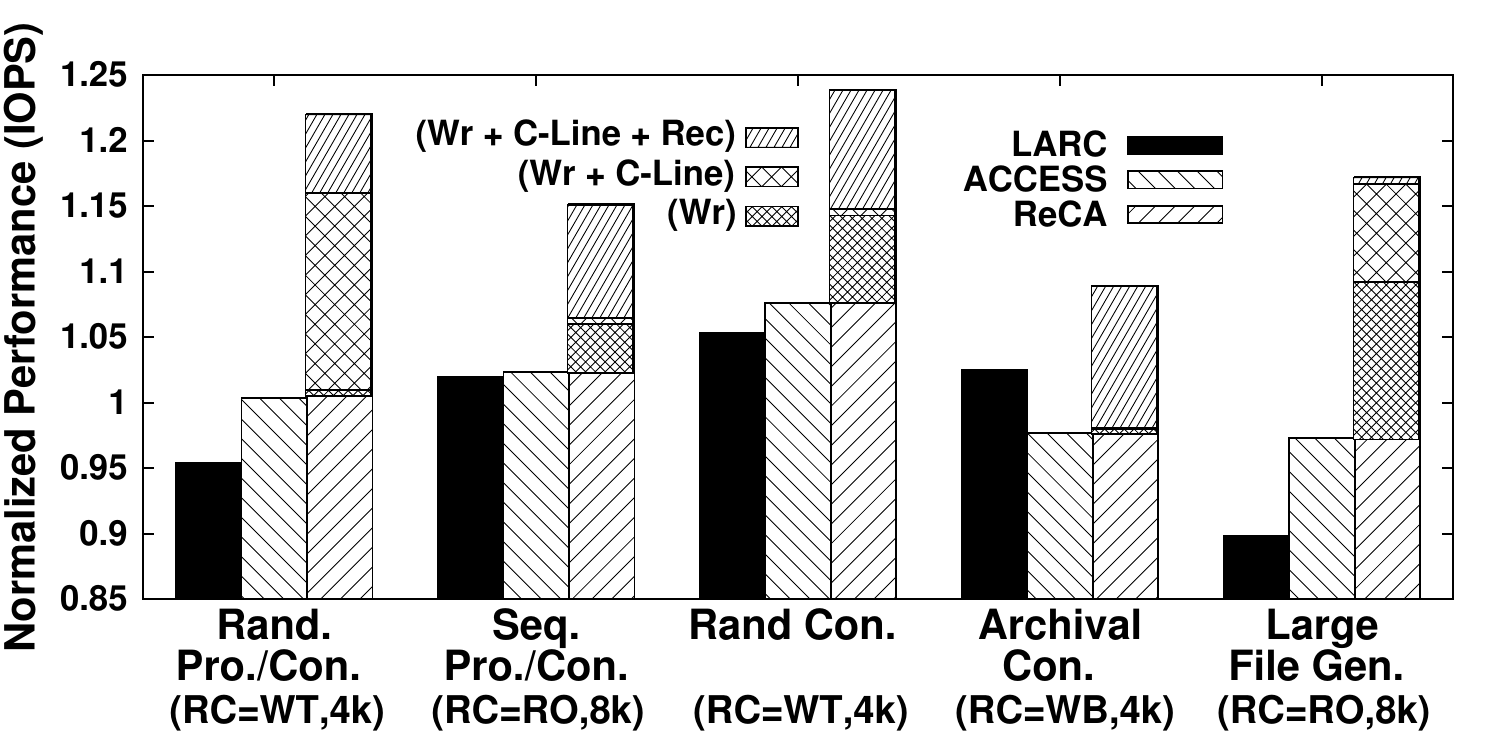}
\vspace{-.23cm}
\caption{Performance of Various Cache Architectures}
\label{fig:last}
\vspace{-.53cm}
\end{figure}

\begin{figure*}
\centering
\includegraphics[width=0.85\textwidth]{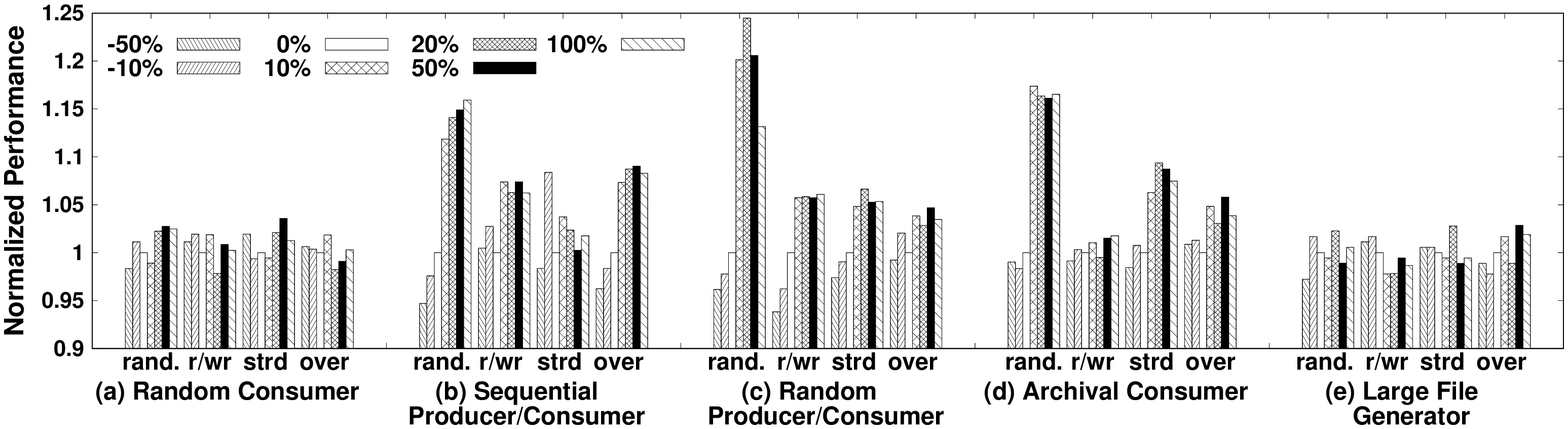}%
\vspace{-.23cm}
\caption{Normalized Performance of Various Cache Parameter Priorities in ReCA}
\label{fig:rec}
\vspace{-.53cm}
\end{figure*}
	
Cache line size, similar to eviction and write policies, is a key factor in determining performance of a caching architecture.
Fig. \ref{fig:cl} depicts the effect of various cache line sizes on average response time of examined workloads
where the cache line size of ReCA is set to fixed values instead of being reconfigurable to show the effect of using various cache line sizes on the performance.
Larger cache line sizes are omitted from Fig. \ref{fig:cl} for the sake of brevity.
Although using smaller cache line sizes enables cache to more accurately select suitable data pages for caching, memory footprint increases as cache line size decreases.
In addition, using larger cache line size improves performance in many workloads (Fig. \ref{fig:cl}a and  Fig. \ref{fig:cl}e).
One of the key observations in Fig. \ref{fig:cl}a is that despite being a random workload, 8KB cache line size outperforms 4KB cache line size in all caching architectures.
Further investigations show that this behavior is also observed by using 8KB as data block size by several applications in \emph{Random Consumers} category.
ReCA has higher performance compared to the other caching architectures in most of the configurations examined in Fig. \ref{fig:cl}.
Reconfiguring cache line size in the runtime enables ReCA to maintain its high performance even in case of a change in workload pattern while other architectures only use a fixed cache line size which will be inefficient in various workloads.

Fig. \ref{fig:last} compares the performance of ReCA with all discussed optimizations along with performance of the other caching architectures, all normalized to the performance of LRU caching policy.
The configuration selected for each workload category is also depicted in Fig. \ref{fig:last}.
As shown in this figure, ReCA outperforms LRU, LARC, and \emph{Access} architectures in all workload categories by employing efficient policies and cache structures.
The performance improvement is higher in random workloads (on average) which are the target categories for using caching architectures.
ReCA improves performance in \emph{Random Producers /Consumers} and \emph{Large File Generators} category by 22\% and 24\%, respectively, compared to LARC which confirms the efficiency of ReCA even in sequential workloads.
The performance gain achieved by ReCA is broken down to the employed optimizations to demonstrate the effect of each parameter on the overall performance.
\emph{Wr}, \emph{C-Line}, and \emph{Rec} denote write policy, cache line size, and reclaiming policy, respectively.
Choosing \emph{write-through} and \emph{read-only} policies in ReCA removes the need for using mirrored configuration since no dirty data page exists in the cache.
Using \emph{read-only} policy also improves SSD lifetime by reducing the number of writes in SSD up to 33\% (not shown in figures).
The number of writes is obtained directly from real SSDs which shows the actual number of writes
	in flash chips including the effect of \emph{Flash Translation Layer} (FTL) optimizations and write amplification factor.

FTL in SSDs tries to reduce the number of actual writes in SSD flash chips required to respond to user write requests.
The average number of extra writes is called \emph{write amplification factor} \cite{Hu:2009:WAA:1534530.1534544}.
By sending less write requests to SSD, ReCA reduces FTL garbage collection overhead that increases performance.
Additionally, FTL will have more clean data blocks to perform optimizations and decrease write amplification factor.
Hence, issuing less writes by ReCA will significantly enhance the SSD lifetime.


\begin{figure}
	\centering
	\includegraphics[scale=0.45]{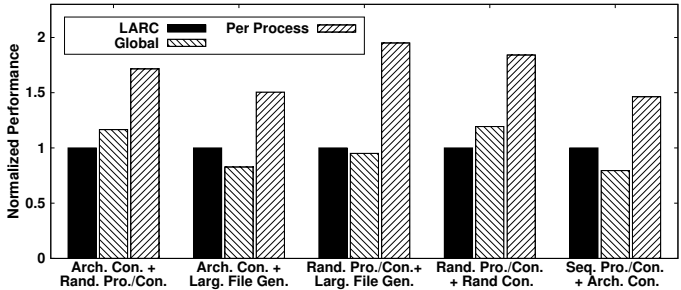}
	\vspace{-.48cm}
	\caption{Performance of \emph{Global} and \emph{per Process} Characterization in ReCA}
	\label{fig:perprocess}
	\vspace{-.6cm}
\end{figure}

\vspace{-.3cm}
\subsection{Parameter Sensitivity}
\label{sec:sensitivity}
\vspace{-.05cm}
The efficiency of selected parameters for prioritizing requests has been investigated in Fig. \ref{fig:rec}.
The experiments have been conducted by both positive and negative values for parameters to fully explore the sensitivity of choosing right value for each parameter.
For each workload category and parameter, a value with highest performance is selected.
ReCA is configured with such values which enables it to simultaneously improve performance, lifetime, and cost (removing the need for mirrored SSDs).

\vspace{-.3cm}
\subsection{Mixed Workloads}
\label{sec:mixedworkload}
\vspace{-.05cm}
ReCA is able to characterize running applications separately to optimize itself based on application requirements.
To show the effectiveness of this technique, applications from different categorizes are run simultaneously \emph{with} and \emph{without} separate characterization.
Fig. \ref{fig:perprocess} depicts the normalized performance of ReCA under many combinations of workloads compared to
LARC.
Configuring cache based on each application requirement can result in more than 2x performance compared to LARC.
Additionally, \emph{per process} caching has almost the same performance gain compared to \emph{global} caching which emphasizes on the importance of separate characterization in ReCA.
Although \emph{Archival consumers} and \emph{Sequential producers/consumers} are both sequential workloads, separating cache configurations results in 80\% performance improvement which demonstrates that the cache configuration can significantly affect the performance in sequential workloads as well as random workloads.

The CPU overhead of \emph{per process} caching is negligible since the CPU overhead is dependent to the number of incoming requests and remains the same in both global and per process characterization.
	However, there is slightly more CPU overhead in per process characterization due to managing extra data structures which is
	negligible in our experiments (less than 2\% CPU usage).
	Although in per process caching characterization data structures are stored per process, they contribute to a small percentage
	of total ReCA memory usage.
	Most of the consumed memory is used for data page mapping in SSD which remains the same in both global and per process characterization.
	The actual increase in memory usage of per process characterization is about 5-10\%.

\begin{figure}
	\centering
	\includegraphics[scale=0.37]{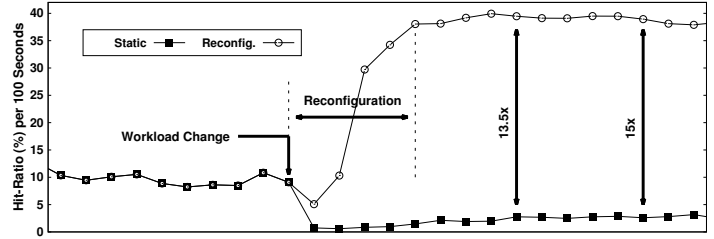}
	\vspace{-.5cm}
	\caption{Hit Ratio of ReCA \emph{with} and \emph{without} Reconfiguration}
	\label{fig:reconfigexp}
	\vspace{-.5cm}
\end{figure}

\vspace{-.3cm}
\subsection{Reconfiguration Overhead}
\label{sec:perfoverheads}
\vspace{-.1cm}
The reconfiguration process consists of \emph{(a)} reconfiguring the data structures and \emph{(b)} eviction of data pages from cache.
The first stage is relatively fast and can be done in less than 10 seconds without interfering with user requests.
The second stage requires issuing I/O requests to both HDD and SSD.
To decrease the impact of reconfiguration process on applications, a limit is put over the number of I/O requests issued for reconfiguration.
To evaluate the efficiency of ReCA reconfiguration and its performance overhead, hit ratio of ReCA \emph{with} and \emph{without} reconfiguration upon changing the workload type is examined.
In both scenarios, workload \emph{fileserver} from \emph{Sequential producer/consumer} category is running and cache configuration is optimized toward its requests.
Workload type is changed into \emph{exchange server} from \emph{Random producer/consumer} category and its impact on the hit ratio is observed.
Fig. \ref{fig:reconfigexp} depicts the hit ratio of ReCA \emph{with} and \emph{without} reconfiguration.
The performance overhead of reconfiguration process is very small and cache reaches its optimal state in less than five minutes of changing workload type.
Although static technique had optimal configuration in the first workload, due to the change in the workload type, it will be completely inefficient in the new workload.
This shortcoming exists in all previous studies in I/O caching \cite{Azor,macss,Hystor,Huang:2016:IFD:2888404.2737832,Santana:2015:AA:2813749.2813763}.

\vspace{-.3cm}
\subsection{Tiering in ReCA}
\label{sec:tieringreca}
\vspace{-.1cm}
Tiering can outperform caching when the workload is steady and contains no sudden changes.
	This is due to the fixed intervals between migrations in tiering (ranges from hours to days), unlike caching which
	can demote a data page from cache or promote a data page to cache for each incoming request.
	Therefore, if one does not expect running applications to change their I/O behavior, tiering is more suitable while if
	sudden changes are likely to happen such as in cloud environments, ReCA will be more efficient to work in caching mode.

\vspace{-.5cm}
\section{Conclusion}
\label{sec:conclusion}
\vspace{-0.15cm}
In this paper, we first demonstrated that solely considering the request type is not effective in prioritizing requests for caching.
Based on this observation, a comprehensive workload characterization is conducted to find an optimal cache configuration for each application type.
In addition, an analysis has been done on filesystem metadata to consider such semantic information in the proposed architecture.
To utilize workload characterization in caching, an online \emph{Reconfigurable Cache Architecture} was proposed for storage systems that monitors incoming application I/Os and reconfigures cache when detecting a change in the running application phase.
In addition, it is revealed that in many applications, mirrored cache requirement can be lifted without any reliability concern while maintaining the performance of application intact.
The lifetime of SSDs used for caching also can be extended in many applications by changing the cache policy to \emph{read-only} without significant degradation in performance.
The experimental results showed that ReCA improves performance up to 24\% (16\% on average) and up to 33\% lifetime improvement compared to previous studies while removing the need for mirrored SSDs in majority of the workloads.

\ifCLASSOPTIONcaptionsoff
  \newpage
\fi



\vspace{-.5cm}
\bibliographystyle{IEEEtran}
\bibliography{ref}
%
%
%

%

\vspace{-.6cm}
\begin{IEEEbiography}[{\includegraphics[width=1in,height=1.25in,clip,keepaspectratio]{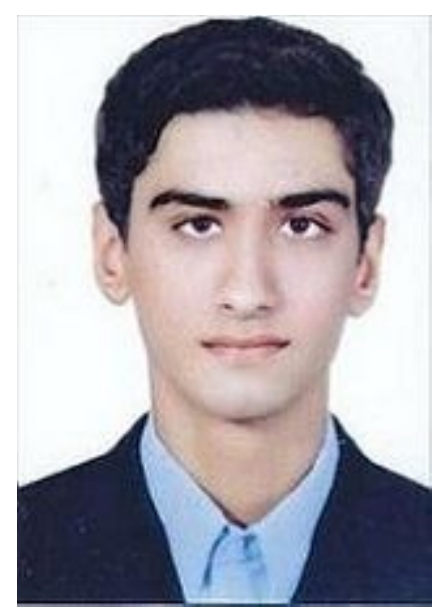}}]{Reza Salkhordeh}
received the B.S. degree in computer engineering from Ferdowsi University of Mashhad in 2011, and M.S. degree in computer
engineering from Sharif University of Technology
(SUT) in 2013. He is currently a Ph.D. candidate at SUT. His research interests include operating systems, solid-state drives, and data
storage systems.
\end{IEEEbiography}
\vfill
\vspace{-.6cm}
\begin{IEEEbiography}[{\includegraphics[width=1in,height=1.25in,clip,keepaspectratio]{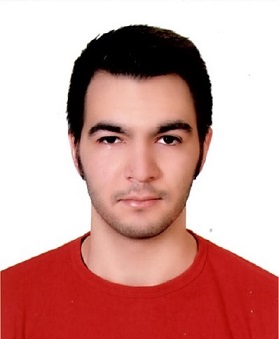}}]{Shahriar Ebrahimi}
Shahriar Ebrahimi received the B.S. degree in computer engineering from Sharif University of Technology (SUT) in 2015. He is currently a M.S. student in computer architecture engineering in SUT. His research interests include storage systems, computer architecture, reconfigurable systems, and application of machine learning in storage systems.
\end{IEEEbiography}
\vfill
\vspace{-.6cm}
\begin{IEEEbiography}[{\includegraphics[width=1in,height=1.25in,clip,keepaspectratio]{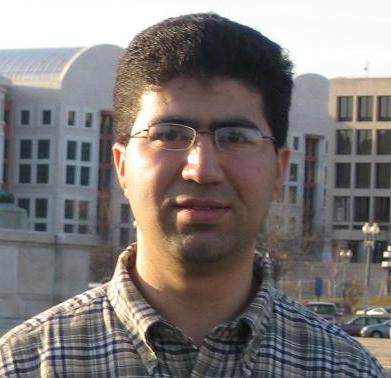}}]{Hossein Asadi}
(M'08, SM'14) received the B.Sc. and M.Sc. degrees in computer engineering from the SUT, Tehran, Iran, in 2000 and 2002, respectively, and the Ph.D. degree in electrical and computer engineering from Northeastern University, Boston, MA, USA, in 2007. 
He was with EMC Corporation, Hopkinton, MA, USA, as a Research Scientist and Senior Hardware Engineer, from 2006 to 2009. From 2002 to 2003, he was a member of the Dependable Systems Laboratory, SUT, where he researched hardware verification techniques. From 2001 to 2002, he was a member of the Sharif Rescue Robots Group. He has been with the Department of Computer Engineering, SUT, since 2009, where he is currently a tenured Associate Professor. He is the Founder and Director of the Data Storage, Networks, and Processing (DSN) Laboratory, Director of Sharif High-Performance Computing Center, the Director of Sharif Information Technology Service Center (ITC), and the President of Sharif ICT Innovation Center. He spent three months in the summer 2015 as a Visiting Professor at the School of Computer and Communication Sciences at the Ecole Poly-technique Federele de Lausanne (EPFL). He is also the co-founder of HPDS corp., designing and fabricating midrange and high-end data storage systems. He has authored and co-authored more than seventy technical papers in reputed journals and conference proceedings. His current research interests include data storage systems and networks, solid-state drives, operating system support for I/O and memory management, and reconfigurable and dependable computing.
Dr. Asadi was a recipient of the Technical Award for the Best Robot Design from the International RoboCup Rescue Competition, organized by AAAI and RoboCup, a recipient of Best Paper Award at the 15th CSI Internation Symposium on Computer Architecture and Digital Systems (CADS), the Distinguished Lecturer Award from SUT in 2010, the Distinguished Researcher Award and the Distinguished Research Institute Award from SUT in 2016, and the Distinguished Technology Award from SUT in 2017. He is also recipient of Extraordinary Ability in Science visa from US Citizenship and Immigration Services in 2008. He has also served as the publication chair of several national and international conferences including CNDS2013, AISP2013, and CSSE2013 during the past four years. Most recently, he has served as a Guest Editor of IEEE Transactions on Computers, a Program Co-Chair of the 18th International Symposium on Computer Architecture \& Digital Systems (CADS2015), and the Program Chair of CSI National Computer Conference (CSICC2017). 
\end{IEEEbiography}
\vfill
%
%
%




\end{document}